\documentclass[fleqn]{lmcs}
\pdfoutput=1

\usepackage{lastpage}
\lmcsdoi{16}{3}{14}
\lmcsheading{}{\pageref{LastPage}}{}{}%
{Sep.~17,~2019}{Aug.~28,~2020}{}

\usepackage[utf8]{inputenc}
\usepackage{graphicx}
\usepackage{color}
\usepackage{amsmath}
\usepackage{amssymb}
\usepackage{amsthm}
\usepackage{underscore}
\usepackage{procalg}
\usepackage[normalem]{ulem}

\newcommand{\straightforwardproof}[1]{}
\newcommand{\technicalproof}[1]{{#1}}
\newcommand{\interestingproof}[1]{{#1}}

\newcommand{\nil}{\ensuremath{\mathalpha{\mathbf{0}}}}
\newcommand{\Acts}{\ensuremath{\mathalpha{\Act{}_{\silent}}}}
\newcommand{\act}[1][\alpha]{\ensuremath{\mathalpha{#1}}}
  \newcommand{\acta}[1][]{\ensuremath{\act[a_{#1}]}}
  
\newcommand{\silent}{\act[\tau]}
\newcommand{\Vars}[1][\mathcal{V}]{\ensuremath{\mathalpha{#1}}}
\newcommand{\var}[1]{\ensuremath{\mathalpha{#1}}}
  \newcommand{\varX}[1][]{\ensuremath{\var{X_{#1}}}}
  \newcommand{\varY}[1][]{\ensuremath{\var{Y_{#1}}}}
  
\newcommand{\FreeVars}[1]{\ensuremath{\mathit{FV}(#1)}}
\newcommand{\PEXP}[1][\mathcal{E}]{\ensuremath{\mathalpha{#1}}}
\newcommand{\pexp}[1]{\ensuremath{\mathalpha{#1}}}
  \newcommand{\pexpE}[1][]{\ensuremath{\pexp{E_{#1}}}}
  \newcommand{\pexpF}[1][]{\ensuremath{\pexp{F_{#1}}}}
  \newcommand{\pexpG}[1][]{\ensuremath{\pexp{G_{#1}}}}
  \newcommand{\pexpH}[1][]{\ensuremath{\pexp{H_{#1}}}}
\newcommand{\CPEXP}[1][\mathcal{P}]{\ensuremath{\mathalpha{#1}}}
\newcommand{\cpexp}[1]{\ensuremath{\mathalpha{#1}}}
  \newcommand{\cpexpP}[1][]{\ensuremath{\cpexp{P_{#1}}}}
  \newcommand{\cpexpQ}[1][]{\ensuremath{\cpexp{Q_{#1}}}}
  \newcommand{\cpexpR}[1][]{\ensuremath{\cpexp{R_{#1}}}}
\newcommand{\XPEXP}[1][\mathcal{P}_{\varX}]{\ensuremath{\mathalpha{#1}}}
\newcommand{\prefsym}[1][\acta]{\ensuremath{\mathalpha{{#1}.}}}
\newcommand{\pref}[2][\acta]{\ensuremath{\prefsym[#1]{#2}}}
\newcommand{\recsym}[1][\varX{}]{\ensuremath{\mathalpha{\mu{#1}.}}}
\newcommand{\rec}[2][\varX{}]{\ensuremath{\recsym[#1]{#2}}}
\newcommand{\subst}[2]{\ensuremath{[{#1}/{#2}]}}

  \newcommand{\statemap}[1][\sigma]{\ensuremath{\mathalpha{#1}}}
  \newcommand{\brelsym}[1][\mathcal{R}]{\ensuremath{#1}}
  \newcommand{\brel}[1][{\brelsym}]{\ensuremath{\mathrel{#1}}}
  \newcommand{\breluptobbisimdsym}[1][\mathcal{R}^{u}]{\ensuremath{#1}}
  \newcommand{\breluptobbisimd}[1][{\breluptobbisimdsym}]{\ensuremath{\mathrel{#1}}}
  \newcommand{\urelsym}[1][\pexpE,\pexpF]{\ensuremath{\mathcal{R}_{#1}}}
  \newcommand{\urel}[1][\pexpE,\pexpF]{\ensuremath{\mathrel{\urelsym[{#1}]}}}

   \newdimen\boxwdplusemdimen
   \def\arrow#1{{
     \boxwdplusemdimen=1em%
     \setbox0=\hbox{$\scriptstyle#1$}%
     \advance\boxwdplusemdimen by \wd0\relax%
     \ifdim\boxwdplusemdimen<16.11119pt%
       \boxwdplusemdimen=16.11119pt%
     \fi%
     \buildrel{#1}\over%
       {\setbox1=\hbox to \boxwdplusemdimen{\rightarrowfill}%
     \ht1=0.3em\relax\box1}%
   }}
   \makeatletter
   \def\twoheadrightarrowfill{$\m@th\smash-\mkern-7mu%
     \cleaders\hbox{$\mkern-2mu\smash-\mkern-2mu$}\hfill
     \mkern-7mu\mathord\twoheadrightarrow$}
   \makeatother
   \def\darrow#1{{
     \boxwdplusemdimen=1em%
     \setbox0=\hbox{$\scriptstyle#1$}%
     \advance\boxwdplusemdimen by \wd0\relax%
     \ifdim\boxwdplusemdimen<16.11119pt%
       \boxwdplusemdimen=16.11119pt%
     \fi%
     \buildrel{#1}\over%
       {\setbox1=\hbox to \boxwdplusemdimen{\twoheadrightarrowfill}%
     \ht1=0.3em\relax\box1}%
   }}
   \def\plarrow#1{{
     \boxwdplusemdimen=1em%
     \setbox0=\hbox{$\scriptstyle#1$}%
     \advance\boxwdplusemdimen by \wd0\relax%
     \ifdim\boxwdplusemdimen<16.11119pt%
       \boxwdplusemdimen=16.11119pt%
     \fi%
     \buildrel{#1}\over%
       {\setbox1=\hbox to \boxwdplusemdimen{\rightarrowfill}%
     \ht1=0.3em\relax\box1^{\scriptstyle +}}%
   }}
  \newcommand{\stepsym}{\ensuremath{\mathalpha{\longrightarrow}}}
  \newcommand{\reachsym}{\stepsym{}^{*}}
  \newcommand{\reach}{\ensuremath{\mathbin{\reachsym}}}
  \newcommand{\ssteps}{\ensuremath{\mathbin{\darrow{}}}}

  \newcommand{\wbbisimd}{%
    \setbox0=\hbox{\kern-.1ex{$\leftrightarrow$}\kern-.1ex}
    \setbox1=\vbox{\hbox{\raise .1ex \box0}\hrule}%
    \ensuremath{\mathrel{\hbox{\kern.1ex\box1\kern.1ex}^{\Delta_3}_{b}}}
  }
  \newcommand{\wbbisimdtwo}{%
    \setbox0=\hbox{\kern-.1ex{$\leftrightarrow$}\kern-.1ex}
    \setbox1=\vbox{\hbox{\raise .1ex \box0}\hrule}%
    \ensuremath{\mathrel{\hbox{\kern.1ex\box1\kern.1ex}^{\Delta_2}_{b}}}
  }
  \newcommand{\onestepbbisimd}{%
    \setbox0=\hbox{\kern-.1ex{$\leftrightarrow$}\kern-.1ex}
    \setbox1=\vbox{\hbox{\raise .1ex \box0}\hrule}%
    \ensuremath{\mathrel{\hbox{\kern.1ex\box1\kern.1ex}^{\Delta_1}_{b}}}
  }
  \newcommand{\bbisimd}{%
    \setbox0=\hbox{\kern-.1ex{$\leftrightarrow$}\kern-.1ex}
    \setbox1=\vbox{\hbox{\raise .1ex \box0}\hrule}%
    \ensuremath{\mathrel{\hbox{\kern.1ex\box1\kern.1ex}^{\Delta}_{b}}}
  }
  \newcommand{\bbisimdX}{%
    \setbox0=\hbox{\kern-.1ex{$\leftrightarrow$}\kern-.1ex}
    \setbox1=\vbox{\hbox{\raise .1ex \box0}\hrule}%
    \ensuremath{\mathrel{\hbox{\kern.1ex\box1\kern.1ex}^{\Delta}_{bX}}}
  }
  \newcommand{\rbbisimd}{%
    \setbox0=\hbox{\kern-.1ex{$\leftrightarrow$}\kern-.1ex}
    \setbox1=\vbox{\hbox{\raise .1ex \box0}\hrule}%
    \ensuremath{\mathrel{\hbox{\kern.1ex\box1\kern.1ex}^{\Delta}_{rb}}}
  }
  \newcommand{\rbbisimdX}{%
    \setbox0=\hbox{\kern-.1ex{$\leftrightarrow$}\kern-.1ex}
    \setbox1=\vbox{\hbox{\raise .1ex \box0}\hrule}%
    \ensuremath{\mathrel{\hbox{\kern.1ex\box1\kern.1ex}^{\Delta}_{rbX}}}
  }
  \newcommand{\rbbisimdzero}{%
    \setbox0=\hbox{\kern-.1ex{$\leftrightarrow$}\kern-.1ex}
    \setbox1=\vbox{\hbox{\raise .1ex \box0}\hrule}%
    \ensuremath{\mathrel{\hbox{\kern.1ex\box1\kern.1ex}^{\Delta_0}_{rb}}}
  }
  
  \newcommand{\bbisimdsym}{\ensuremath{\mathalpha{\bbisimd}}}

  \newcommand{\N}{\ensuremath{\mathalpha{\omega}}}

  \newcommand{\relcomp}{\ensuremath{\mathbin{;}}}
\newcommand{\opt}[1]{\mbox{\tiny\rm(}#1\mbox{\tiny\rm)}} 
\newcommand{\plat}[1]{\raisebox{0pt}[0pt][0pt]{#1}}      


\begin{document}

\title{Rooted Divergence-Preserving Branching Bisimilarity is a Congruence}

\author[R.J. van Glabbeek]{Rob van Glabbeek\rsuper{a}}
\address{\lsuper{a}Data61, CSIRO, Sydney, Australia\newline
              School of Computer Science and Engineering,
              University of New South Wales, Sydney, Australia}
\email{rvg@cs.stanford.edu}
            
\author[B. Luttik]{Bas Luttik\rsuper{b}}
\address{\lsuper{b}Department of Mathematics and Computer Science,
              Eindhoven University of Technology,
              The Netherlands}
            \email{s.p.luttik@tue.nl}

\author[L. Spaninks]{Linda Spaninks\rsuper{b}}

\begin{abstract}\noindent
  We prove that rooted divergence-preserving branching bisimilarity is
  a congruence for the process specification language consisting of
  nil, action prefix, choice, and the recursion construct.
\end{abstract}

\maketitle

\section{Introduction}\label{sect:introduction}

Branching bisimilarity \cite{GW96} is a behavioural equivalence on
processes that is compatible with abstraction from internal activity,
while at the same time preserving the branching structure of processes in a
strong sense \cite{vG01a}.
Branching bisimilarity abstracts to a large extent from
\emph{divergence} (i.e., infinite internal activity). For instance, it
identifies a process, say $P$, that may perform some internal activity
after which it returns to its initial state (i.e., $P$ has a
$\silent$-loop) with a process, say $P'$, that admits the same
behaviour as $P$ except that it cannot perform the internal activity
leading to the initial state (i.e., $P'$ is $P$ without the
$\silent$-loop).

In situations where fairness principles apply, abstraction from
divergence is often desirable. But there are circumstances in which
abstraction from divergence is undesirable: A behavioural equivalence that
abstracts from divergence is not compatible with any temporal logic
featuring an \emph{eventually} modality: for any desired state that
$P'$ will eventually reach, the mentioned internal activity of $P$ may
be performed forever, and thus prevent $P$ from reaching this desired
state. It is also generally not compatible with a process-algebraic
priority operator (cf.\ \cite[pp.\ 130--132]{Vaa90}) or sequencing operator
\cite{BLY17}. Since a divergence may be exploited to simulate
recursively enumerable branching in a computable transition system
\cite{Phi93}, a divergence-insensitive behavioural equivalence may be
considered too coarse for a theory that integrates computability and
concurrency \cite{BLT13}.  Preservation of divergence is widely
considered an important correctness criterion when studying the
relative expressiveness of process calculi \cite{Gor10,Xu12,Fu16}.

The notion of \emph{branching bisimilarity with explicit divergence},
also stemming from \cite{GW96}, is a suitable refinement of branching
bisimilarity that is compatible with the well-known branching-time
temporal logic CTL$^{*}$ without the nexttime operator $X$ (which is
known to be incompatible with abstraction from internal activity).
In fact, in \cite{GLT09b} we have proved that it is the coarsest
semantic equivalence on labelled transition systems with silent
moves that is a congruence for parallel composition (as found in
process algebras like CCS, CSP or ACP) and only equates processes
satisfying the same CTL$^{*}_{-X}$ formulas.
In \cite{BLT13}, for stylistic reasons, \emph{branching
  bisimilarity with explicit divergence} was named
\emph{divergence-preserving branching bisimilarity}; we shall
henceforth use this term.

Divergence-preserving branching bisimilarity is the finest behavioural
equivalence in the linear time -- branching time spectrum
\cite{Gla93a}. It is the principal behavioural equivalence underlying
the theory of executability \cite{BLT12,BLT13,LY15,LY16}. Reduction
modulo divergence-preserving branching bisimilarity is a part of
methods for formal verification and analysis of the behaviour of
systems \cite{MW14,WE13,PW17,ZSWYZM16}. In \cite{EKW16} a game-based
characterisation of divergence-preserving branching bisimilarity is
presented.

Processes are usually specified in some process specification
language. For compositional reasoning it is then important that the
behavioural equivalence used is a congruence with respect to the
constructs of that language. Following Milner \cite{Mil89}, we
consider the language \emph{basic CCS with recursion}, i.e., the language consisting of
$\nil$, action prefix, and choice, extended with the recursion
construct $\rec[\varX]{\_}$; this language precisely allows the
specification of finite-state behaviours. As for other weak
behavioural equivalences, divergence-preserving branching bisimilarity is not a congruence for
that language; in fact, it is not a congruence for any language that
includes choice. The goal of this paper is to prove that adding the
usual root condition suffices to obtain a congruence---and, in fact,
the coarsest congruence---for the language under consideration that is
included in divergence-preserving branching
bisimilarity. Interestingly, the root condition is not only necessary
to get a congruence for choice, but also for recursion: $\pref[\tau]X$
is divergence-preserving branching bisimilar to $X$, yet $\rec[\varX]{\pref[\tau]\varX}$ diverges whereas $\rec[\varX]{\varX}$ does not.

Recently, a congruence format was proposed for (rooted)
divergence-preserving branching bisimilarity \cite{FvGL2019}.
The operational rules for action prefix and choice
are in this format. Unfortunately, however, this format does not
support the recursion construct $\rec[\varX]{\_}$. Interestingly, as
far as we know, the recursion construct has not been covered at all in
the rich literature on congruence formats, with the recent exception of
\cite{vG17}. (The article \cite{vG17} differentiates between
\emph{lean} and \emph{full} congruences for recursion; in this article
we consider the full congruence.)

The congruence result obtained in this paper should serve as a
stepping stone towards
a complete axiomatisation of divergence-preserving branching
bisimilarity for basic CCS with recursion. Such work, inspired by
Milner's complete axiomatisation of weak bisimilarity \cite{Mil89},
would combine the adaptations of \cite{Gla93} to branching
bisimilarity, and of \cite{LDH05} to several divergence-sensitive
variants of weak bisimilarity.

We originally thought that congruence for recursion could be obtained
in the same spirit as Milner's ingenious proof in \cite{Mil89h} for
strong bisimilarity, which cleverly makes use of an up-to
technique. The proofs for weak and branching bisimilarity essentially
reuse this idea \cite{Mil89h,Gla93}, but require the use of a
weak step in the antecedent of the transfer condition. We were not
able to generalise the idea to divergence-preserving branching
bisimilarity until we included the root condition in the up-to technique.

We believe that the proofs of Corollaries~\ref{cor:liftuprec} and \ref{cor:liftuprec-inf},
Propositions~\ref{prop:openbis} and
\ref{prop:uptorelatedimpliesrbbisimd}, and Lemma~\ref{lem:ureldiv}
contain novel twists. Although the other proofs are either routine or
adaptations of the ones in \cite{Mil89h}, we have included them for
the convenience of the reader.

In this paper we do not study the CCS constructs for parallel
composition, restriction and relabelling. However, combining our
results with those of \cite{FvGL2019} yields a congruence result for full CCS
with the proviso that parallel composition, restriction and
relabelling are not allowed in the scope of a recursion.
This spans most practical applications.
The method employed in this paper does not generalise to obtain a
congruence result for full CCS, featuring parallel composition in
the scope of recursion. Although we conjecture that such a
congruence result holds, proving it remains an open problem.

\section{Rooted divergence-preserving branching bisimilarity} \label{sec:bb}

Let $\Act$ be a non-empty set of \emph{actions}, and let $\silent$ be
a special action not in $\Act$. Let $\Acts=\Act\cup\{\silent\}$.
Furthermore, let $\Vars$ be a set of \emph{recursion variables}.
The set of \emph{process expressions} $\PEXP$ is generated by the
following grammar:
\begin{equation*}
  \pexpE ::= \nil\ \mid\ \varX\ \mid\ 
                   \pref[\act]{\pexpE}\ \mid\ \rec[\varX]{\pexpE}\ \mid\ 
                   \pexpE\altc\pexpE
     \qquad (\act\in\Acts,\ \varX\in\Vars)
\enskip.
\end{equation*}

An occurrence of a recursion variable $\varX$ in a process expression
$\pexpE$ is \emph{bound} if it is in the scope of a $\recsym[\varX]{\_}$,
and otherwise it is \emph{free}. We denote by $\FreeVars{\pexpE}$ the
set of variables with a free occurrence in $\pexpE$. If
$\vec{\varX}=\varX[0],\dots,\varX[n]$ is a sequence of variables, and
$\vec{\pexpF}=\pexpF[0],\dots,\pexpF[n]$ is a sequence of process
expressions of the same length, then we write
$\pexpE\subst{\vec{\pexpF}}{\vec{\varX}}$ for the process expression
obtained from $\pexpE$ by replacing all free occurrences of $\varX[i]$
in $\pexpE$ by $\pexpF[i]$ ($i=0,\dots,n$), applying
$\alpha$-conversion to $\pexpE$ if necessary to avoid capture.

On $\PEXP$ we define an $\Acts$-labelled transition relation
${\step{}}\subseteq\PEXP\times\Acts\times\PEXP$ as the least ternary
relation satisfying the following rules for all $\act\in\Acts$,
$\varX\in\Vars$, and process expressions $\pexpE$, $\pexpE'$, $\pexpF$
and $\pexpF'$:

\begin{center}
\begin{osrules}
  \osrule{}{\pref[\act]{\pexpE}\step{\act}\pexpE}
    \label{osrule:pref}
\quad
   \osrule{\pexpE\subst{\rec[\varX]{\pexpE}}{\varX} \step{\act} \pexpE'}%
              {\rec[\varX]{\pexpE}\step{\act}\pexpE'}
       \label{osrule:rec}
\quad
   \osrule{\pexpE\step{\act}\pexpE'}{\pexpE\altc\pexpF\step{\act}\pexpE'}
  \label{osrule:altcleft}
\quad
   \osrule{\pexpF\step{\act}\pexpF'}{\pexpE\altc\pexpF\step{\act}\pexpF'}
  \label{osrule:altcright}
\end{osrules}
\end{center}

We write $\pexpE\step{\act}\pexpE'$ for
$(\pexpE,\act,\pexpE')\in\stepsym$ (as we already did in the rules
  above) and we abbreviate the 
statement `$\pexpE\step{\act}\pexpE'$ or ($\act{}=\silent$ and
$\pexpE=\pexpE'$)' by
\plat{$\pexpE\step{\opt{\act}}\pexpE'$}. Furthermore, we write
$\ssteps{}$ for the reflexive-transitive closure of $\step{\silent}$,
i.e., $\pexpE\ssteps{}\pexpE'$ if there exist
$\pexpE[0],\pexpE[1],\dots,\pexpE[n]\in\PEXP$ such that
  $\pexpE=\pexpE[0]\step{\silent}\pexpE[1]\step{\silent}\cdots\step{\silent}\pexpE[n]=\pexpE'$.

  A process expression is \emph{closed} if it contains no free
  occurrences of recursion variables; we denote by $\CPEXP$ the subset
  of $\PEXP$ consisting of all closed process expressions. It is easy
  to check that if $\cpexpP$ is a closed process expression and
  $\cpexpP\step{\act}\pexpE$, then $\pexpE$ is a closed process
  expression too. Hence, the transition relation restricts in a
  natural way to closed process expressions, and thus associates with every closed process
expression a behaviour. We proceed to define when two process
expressions may be considered to represent the same
behaviour.

\begin{defi} \label{def:bbisimd}
  A symmetric binary relation $\brelsym$ on $\CPEXP$ is a
  \emph{branching bisimulation} if it satisfies the following
  condition for all $\cpexpP,\cpexpQ\in\CPEXP$ and $\act{}\in\Acts$:
  \begin{enumerate}[align=right,widest=(D)]
  \renewcommand{\theenumi}{T}
  \renewcommand{\labelenumi}{(\theenumi)}
  \item \label{cnd:stepsim}
    if $\cpexpP\brel\cpexpQ$ and $\cpexpP\step{\act}\cpexpP'$ for
    some closed process expression $\cpexpP'$,
    then there exist closed process expressions $\cpexpQ'$ and $\cpexpQ''$
    such that
      \plat{$\cpexpQ\ssteps{}\cpexpQ''\step{\opt{\act}}\cpexpQ'$},
      $\cpexpP\brel\cpexpQ''$
    and
      $\cpexpP'\brel\cpexpQ'$.
  \end{enumerate}

  We say that a branching bisimulation $\brelsym$ \emph{preserves
    (internal) divergence}
  if
  \begin{enumerate}[align=right,widest=(D)]
  \renewcommand{\labelenumi}{(\theenumi)}
  \renewcommand{\theenumi}{D}
  \item \label{cnd:divsim}
     if $\cpexpP\brel\cpexpQ$ and there is an infinite sequence of
     closed process expressions
         $(\cpexpP[k])_{k\in\N}$
       such that
          $\cpexpP=\cpexpP[0]$,
          $\cpexpP[k]\step{\silent}\cpexpP[k+1]$
       and
          $\cpexpP[k]\brel\cpexpQ$ for all $k\in\N$,
      then there is an infinite sequence of closed process expressions
        $(\cpexpQ[\ell])_{\ell\in\N}$
       such that
          $\cpexpQ=\cpexpQ[0]$,
          $\cpexpQ[\ell]\step{\silent}\cpexpQ[\ell+1]$
       and
          $\cpexpP[k]\brel\cpexpQ[\ell]$ for all $k,\ell\in\N$.
  \end{enumerate}
  We write $\cpexpP\bbisimd\cpexpQ$ if there exists a
  divergence-preserving branching bisimulation $\brelsym$ such that
  $\cpexpP\brel\cpexpQ$. The relation $\bbisimd$ was introduced in
  \cite{GW96} under the name branching bisimilarity with explicit
  divergence and is here referred to as divergence-preserving
  branching bisimilarity.
\end{defi}

The relation $\bbisimd$ was studied in detail in \cite{GLT09a}; we
recap some of the facts established \textit{ibidem}.

First, the relation $\bbisimd$ is an equivalence relation.
Second, the relation $\bbisimd$ satisfies the condition (\ref{cnd:stepsim}), with
the following generalisation as a straightforward consequence.
\begin{lem}\label{lem:longstepsim}
  Let $\cpexpP$ and $\cpexpQ$ be closed process expressions.
  If $\cpexpP\bbisimd\cpexpQ$ and
  $\cpexpP\ssteps{}\cpexpP''\step{\act}\cpexpP'$ for some closed
  process expressions $\cpexpP'$ and $\cpexpP''$, then there exist
  closed process expressions $\cpexpQ'$ and $\cpexpQ''$ such that
  $\cpexpQ\ssteps{}\cpexpQ''\step{\act}\cpexpQ'$,
  $\cpexpP''\bbisimd\cpexpQ''$ and $\cpexpP'\bbisimd\cpexpQ'$.
\end{lem}
\straightforwardproof{%
\begin{proof}
  Let $\cpexpP[0],\dots,\cpexpP[n]$ be closed process expressions such
  that
     $\cpexpP=\cpexpP[0]$, $\cpexpP''=\cpexpP[n]$, and
     $\cpexpP[i]\step{\silent}\cpexpP[i+1]$ for all $0\leq i <n$.
  Then, since $\bbisimd$ satisfies condition~(\ref{cnd:stepsim}) of
  Definition~\ref{def:bbisimd}, it follows with a straightforward
  induction on $n$ that there exist closed process expressions
  $\cpexpQ[0],\dots,\cpexpQ[n]$ such that
     $\cpexpQ=\cpexpQ[0]$,
     $\cpexpQ[i]\ssteps\cpexpQ[i+1]$ and
     $\cpexpP[i+1]\bbisimd\cpexpQ[i+1]$ for all $0\leq i < n$.
   Furthermore, since $\cpexpP''=\cpexpP[n]\bbisimd\cpexpQ[n]$ and
   $\cpexpP''\step{\act}\cpexpP'$, it follows from
   condition~\ref{cnd:stepsim} that there exist $\cpexpQ''$ and
   $\cpexpQ'$ such that
   $\cpexpQ[n]\ssteps{}\cpexpQ''\step{\act}\cpexpQ'$,
   $\cpexpP''\bbisimd\cpexpQ''$ and $\cpexpP'\bbisimd\cpexpQ'$.
  Since, clearly, $\cpexpQ\ssteps\cpexpQ''$, the proof is complete.
\end{proof}}
Third, $\bbisimd$ also satisfies (\ref{cnd:divsim}). In \cite{GLT09a} several alternative definitions of divergence
preservation are studied, which, in the end, all give rise to the same
notion of divergence-preserving branching bisimilarity. In particular,
the following alternative relational characterisations will be useful
tools in the remainder.
\begin{prop} \label{prop:divsimalternatives}
  Let $\cpexpP$ and $\cpexpQ$ be closed process expressions. Then
\begin{itemize}
\item
  $\cpexpP\bbisimd\cpexpQ$ if, and only if,
  $\cpexpP$ and $\cpexpQ$ are related by some branching bisimulation
  $\brelsym{}$ satisfying
  \begin{enumerate}[align=right,widest=(D''),topsep=1ex]
  \renewcommand{\labelenumi}{(\theenumi)}
  \renewcommand{\theenumi}{D$'$}
  \item \label{cnd:divsimmap}
     if $\cpexpP\brel\cpexpQ$ and there is an infinite sequence of
     closed process expressions
         $(\cpexpP[k])_{k\in\N}$
       such that
          $\cpexpP=\cpexpP[0]$ and
          $\cpexpP[k]\step{\silent}\cpexpP[k+1]$,
      then there is an infinite sequence of closed process expressions
        $(\cpexpQ[\ell])_{\ell\in\N}$ and
      a mapping $\statemap: {\N\rightarrow\N}$ 
      such that
          $\cpexpQ=\cpexpQ[0]$,
          $\cpexpQ[\ell]\step{\silent}\cpexpQ[\ell+1]$
      and
         $\cpexpP[\statemap(\ell)]\brel\cpexpQ[\ell]$ for all
         $\ell\in\N$; and
  \end{enumerate}
\item
  $\cpexpP\bbisimd\cpexpQ$ if, and only if,
  $\cpexpP$ and $\cpexpQ$ are related by some branching bisimulation
  $\brelsym{}$ satisfying
  \begin{enumerate}[align=right,widest=(D'') ,topsep=1ex]
  \renewcommand{\labelenumi}{(\theenumi)}
  \renewcommand{\theenumi}{D$''$}
  \item \label{cnd:divsimshort}
     if $\cpexpP\brel\cpexpQ$ and there is an infinite sequence of
     closed process expressions
         $(\cpexpP[k])_{k\in\N}$
       such that
          $\cpexpP=\cpexpP[0]$ and
          $\cpexpP[k]\step{\silent}\cpexpP[k+1]$,
      then there exists a closed process expression
        $\cpexpQ'$
      such that
          $\cpexpQ\step{\silent}\cpexpQ'$
      and
         $\cpexpP[k]\brel\cpexpQ'$ for some $k\in\N$.
  \end{enumerate}
\end{itemize}
  Moreover, $\bbisimd$ itself satisfies (\ref{cnd:divsimmap}) and (\ref{cnd:divsimshort}).
\end{prop}
\technicalproof{%
\begin{proof}
  See \cite{GLT09a}; condition (\ref{cnd:divsimmap}) is (D3) and condition (\ref{cnd:divsimshort}) is (D2).
\end{proof}%
}

And finally, it was proved in \cite{GLT09a} that $\bbisimd$ satisfies the
following so-called stuttering property.

\begin{prop}\label{prop:bbisimdstuttering}
  Let $\cpexpP$ be a closed process expression and let
  $\cpexpQ[0],\dots,\cpexpQ[k]$ be closed process expressions such
  that $\cpexpQ[0]\step{\silent}\cdots\step{\silent}\cpexpQ[k]$.  If
  $\cpexpP\bbisimd\cpexpQ[0]$ and $\cpexpP\bbisimd\cpexpQ[k]$, then
  $\cpexpP\bbisimd\cpexpQ[i]$ for all $0\leq i \leq k$.
\end{prop}

As for all variants of bisimilarity that take some form of abstraction from internal activity into account, the relation $\bbisimd$ is not compatible with
$\altc$ 
($\nil\bbisimd\pref[\silent]\nil$ but
 $\nil\altc\pref[\acta]\nil\not\bbisimd\pref[\silent]\nil\altc\pref[\acta]\nil$), and hence not a congruence for the language we
are considering. In contrast to its
divergence-insensitive variant, divergence-preserving branching
bisimilarity is not compatible with the recursion construct
either, as we will argue below. Similarly as for the
divergence-insensitive variant of branching
bisimilarity, it suffices to add a \emph{root condition} to obtain the
coarsest congruence for our language that is contained in $\bbisimd$,
as we shall prove in the remainder of this paper.

\begin{defi} \label{def:rbbisimd}
  Let $\cpexpP$ and $\cpexpQ$ be closed process expressions. We say
  that $\cpexpP$ and $\cpexpQ$ are \emph{rooted divergence-preserving
  branching bisimilar} (notation: $\cpexpP\rbbisimd\cpexpQ$) if for all
  $\act{}\in\Acts$ the following holds:
  \begin{enumerate}[align=right,widest=(R2)]
  \renewcommand{\labelenumi}{(\theenumi)}
  \renewcommand{\theenumi}{R\arabic{enumi}}
   \item\label{cnd:rootsiml}
    if $\cpexpP\step{\act{}}\cpexpP'$, then there exists a $\cpexpQ'$
    such that $\cpexpQ\step{\act{}}\cpexpQ'$ and
    $\cpexpP'\bbisimd\cpexpQ'$; and
  \item\label{cnd:rootsimr}
    if $\cpexpQ\step{\act{}}\cpexpQ'$, then there exists a
    $\cpexpP'$ such that $\cpexpP\step{\act{}}\cpexpP'$ and $\cpexpP'\bbisimd\cpexpQ'$.
  \end{enumerate}
\end{defi}

\noindent
The following proposition is a straightforward consequence of the fact that $\bbisimd$ is an equivalence.
\begin{prop}\label{prop:equivalence}
  The relation $\rbbisimd$ is an equivalence relation on $\CPEXP$.\qed
\end{prop}
Moreover, it is immediate that ${\rbbisimd}\subseteq{\bbisimd}$.
It is well known, and follows immediately from the definition,
that $P\rbbisimd Q$ iff $P+f.0\bbisimd Q+f.0$ for a fresh action $f$,
not occurring in $P$ or $Q$. Using this, the problem of checking rooted divergence-preserving
branching bisimilarity reduces trivially to that of checking divergence-preserving branching bisimilarity.

We have defined the notions of $\bbisimd$ and $\rbbisimd$ on closed
process expressions because those are thought of as directly
representing behaviour. Due to the presence of the binding construct
$\recsym[\varX]{\_}$ it is, however, necessary to lift these notions to
expressions with free variables even if the goal is simply to
establish behavioural equivalence of closed process expressions.

\begin{defi} \label{def:rbbisimdopen}
  Let $\pexpE$ and $\pexpF$ be process expressions, and let the
  sequence $\vec{\varX}$
  of variables at least include all the variables with a free
  occurrence in $\pexpE$ or $\pexpF$. We write $\pexpE\rbbisimd\pexpF$
  ($\pexpE\bbisimd\pexpF$) if
    $\pexpE\subst{\vec{\cpexpP}}{\vec{\varX}}\rbbisimd\pexpF\subst{\vec{\cpexpP}}{\vec{\varX}}$
  ($\pexpE\subst{\vec{\cpexpP}}{\vec{\varX}}\bbisimd\pexpF\subst{\vec{\cpexpP}}{\vec{\varX}}$)
  for every sequence of closed process expressions
    $\vec{\cpexpP}$
  of the same length as $\vec{\varX}$.
\end{defi}

It is clear from the definition above that, since $\bbisimd$ and $\rbbisimd$ are
equivalence relations on $\CPEXP$, their lifted versions are equivalence
relations on $\PEXP$. Note that $\bbisimd$ is not compatible with
  the recursion construct: we have that $X\bbisimd \pref[\silent]{X}$,
  whereas $\rec[\varX]{X}\not\bbisimd\rec[\varX]{\pref[\silent]{X}}$.
We shall prove that its rooted variant $\rbbisimd$ is, however, compatible with all the
constructs of the syntax, i.e.,\ if $\pexpE\rbbisimd\pexpF$, then
$\pref[\act]{\pexpE}\rbbisimd\pref[\act]{\pexpF}$ for all
$\act\in\Acts$, $\rec[\varX]{\pexpE}\rbbisimd\rec[\varX]{\pexpF}$ for
all $\varX\in\Vars$,
$\pexpE\altc\pexpH\rbbisimd\pexpF\altc\pexpH$ and
$\pexpH\altc\pexpE\rbbisimd\pexpH\altc\pexpF$ for all process
expressions $\pexpH$. To prove that $\rbbisimd$ is compatible with
$\prefsym[\act]{}$ and $\altc$ is straightforward, but for
$\recsym[\varX]{\_}$ this is considerably more work.

\section{The congruence proof}

  Our proof that \plat{$\rbbisimd$} is compatible with $\recsym[\varX]{\_}$
  relies on the following observation:
  If $\vec{\varY}$ is some sequence of variables and $\vec{\cpexpP}$
  is a sequence of closed terms of the same length, then, on the one
  hand, $\pexpE\mathrel{\rbbisimd}\pexpF$ implies
  $\pexpE\subst{\vec{\cpexpP}}{\vec{\varY}}\rbbisimd\pexpF\subst{\vec{\cpexpP}}{\vec{\varY}}$
  by the definition of $\rbbisimd$ on $\PEXP$, 
  and, on the other hand, if $\varX$ does not occur in $\vec{\varY}$,
  then from $\rec[\varX]{\pexpE\subst{\vec{\cpexpP}}{\vec{\varY}}}\rbbisimd\rec[\varX]{\pexpF\subst{\vec{\cpexpP}}{\vec{\varY}}}$
  it follows that $(\rec[\varX]{\pexpE})\subst{\vec{\cpexpP}}{\vec{\varY}}\rbbisimd(\rec[\varX]{\pexpF})\subst{\vec{\cpexpP}}{\vec{\varY}}$
    by the definition of substitution.
  Therefore, as formalised in the proof of Proposition~\ref{prop:compatibility}, it is enough to establish that $\pexpE\rbbisimd\pexpF$
  implies $\rec[\varX]{\pexpE}\rbbisimd\rec[\varX]{\pexpF}$ in the
  special case that $\pexpE$ and $\pexpF$ are process expressions
  with no other free variables than $\varX$; such process expressions
  will be called $\varX$-closed.

  The rest of this section is organised as follows.

  We shall first characterise, in Section~\ref{subsec:Xclosed}, the
  relation $\rbbisimd$ on $\varX$-closed process expressions in terms
  of the transition relation on $\varX$-closed process
  expressions.

  Then, in Section~\ref{subsec:rdpbbupto}, we shall
  present a suitable notion of rooted divergence-preserving branching
  bisimulation up to $\rbbisimd$, and we shall prove that every
  pair of rooted divergence-preserving branching bisimilar
  $\varX$-closed process expressions $(\pexpE,\pexpF)$ gives rise to a
  relation $\breluptobbisimdsym$ of which we can show that it is a
  rooted divergence-preserving branching bisimulation up to
  $\rbbisimd$. The relation $\breluptobbisimdsym$ will be defined in
  such a way that it relates $\rec[\varX]{\pexpE}$ and
  $\rec[\varX]{\pexpF}$ and thus allows us to conclude that these
  process expressions are rooted divergence-preserving bisimilar.

  In Section~\ref{subsec:main}, we shall then put the pieces
  together and prove $\rbbisimd$ is the coarsest congruence contained
  in $\bbisimd$ for basic CCS with recursion.

\subsection[Bisimilarity on X-closed process expressions]
{\texorpdfstring{$\rbbisimd$}{Bisimilarity} on \texorpdfstring{$\varX$}{X}-closed process expressions}
\label{subsec:Xclosed}

We say that a process expression $\pexpE$ is \emph{$\varX$-closed} if
$\FreeVars{\pexpE}\subseteq\{\varX\}$; the set of all $\varX$-closed
process expressions is denoted by $\XPEXP$. Note that if $\pexpE$ is
$\varX$-closed and $\pexpE\step{\act}\pexpE'$, then $\pexpE'$ is
$\varX$-closed too, and so the $\Actt$-labelled transition relation restricts in a
  natural way to $\varX$-closed process expressions.

\begin{defi}\label{df:exposed}
  We define when $\varX$ is \emph{exposed} in a (not necessarily
  $\varX$-closed) process expression $\pexpE$ by induction on the
  structure of $\pexpE$:
  \begin{enumerate}
  \renewcommand{\theenumi}{\roman{enumi}}
  \item if $\pexpE=\varX$, then $\varX$ is exposed in $\pexpE$;
  \item if $\pexpE=\rec[\varY]\pexpE'$, $\varY$ is a recursion
    variable distinct from $\varX$ and $\varX$ is exposed in
    $\pexpE'$, then $\varX$ is exposed in $\pexpE$;
  \item if $\pexpE=\pexpE[1]\altc\pexpE[2]$ and $\varX$ is exposed in
    $\pexpE[1]$  or $\pexpE[2]$, then $\varX$ is exposed in $\pexpE$.
  \end{enumerate}
\end{defi}
\noindent
Note that the variable $\varX$ is exposed in $\pexpE$ if, and only if, $\pexpE$ has
an unguarded occurrence of $\varX$ in the sense of \cite{Mil89}.

We establish a relationship between the transitions of a closed
process expression $\pexpE\subst{\cpexpP}{\varX}$ that is obtained by
substituting a closed process expression $\cpexpP$ for the variable
$\varX$ in an $\varX$-closed process expression $\pexpE$, and the
transitions of $\pexpE$ and $\cpexpP$.
\begin{lem}\label{lem:opensteps}
  Let $\pexpE$ be an $\varX$-closed process expression, and let
  $\cpexpP$ be a closed process expression.
  \begin{enumerate}
  \item\label{pushdown}
    If $\pexpE\step{\act}\pexpE'$, then
      $\pexpE\subst{\cpexpP}{\varX}\step{\act}\pexpE'\subst{\cpexpP}{\varX}$,
    and if $\varX$ is exposed in $\pexpE$ and
    $\cpexpP\step{\act{}}\cpexpP'$, then $\pexpE\subst{\cpexpP}{\varX}\step{\act{}}\cpexpP'$.
  \item\label{liftup}
    If $\pexpE\subst{\cpexpP}{\varX}\step{\act}\cpexpP'$ for some (closed)
    process expression $\cpexpP'$\!, then either there exists an
    $\varX$-closed process expression $\pexpE'$ such that
    $\pexpE\step{\act}\pexpE'$ and
    $\cpexpP'=\pexpE'\subst{\cpexpP}{\varX}$, or $\varX$ is exposed in
    $\pexpE$, $\cpexpP\step{\act{}}\cpexpP'$ and every derivation
    of $\pexpE\subst{\cpexpP}{\varX}\step{\act}\cpexpP'$ has a
    derivation of $\cpexpP\step{\act}\cpexpP'$ as a subderivation.
\end{enumerate}
\end{lem}
\technicalproof{%
\begin{proof}
  Statement~\ref{pushdown} of the lemma is established with
  straightforward inductions on a derivation of
  $\pexpE\step{\act}\pexpE'$ and on the structure of $\pexpE$.

  We proceed to establish by induction on a derivation of
    $\pexpE\subst{\cpexpP}{\varX}\step{\act}\cpexpP'$
  that
    there exists an $\varX$-closed process expression
    $\pexpE'$ such that $\pexpE\step{\act}\pexpE'$ and
    $\cpexpP'=\pexpE'\subst{\cpexpP}{\varX}$, or $\varX$ is exposed in
    $\pexpE$, $\cpexpP\step{\act{}}\cpexpP'$ and a derivation of
    $\cpexpP\step{\act{}}\cpexpP'$ appears as a subderivation of the
    considered derivation of
    $\pexpE\subst{\cpexpP}{\varX}\step{\act}\cpexpP'$.
  This implies statement~\ref{liftup}.

  We distinguish cases according to the structure of $\pexpE$:

\begin{itemize}
\item
   Clearly, $\pexpE$ cannot be $\nil$, for if $\pexpE=\nil$, then
  $\pexpE\subst{\cpexpP}{\varX}=\nil$, and $\nil$ does not admit any
  transitions.

\item
  If $\pexpE=\varX$, then $\varX$ is exposed in $\pexpE$ and
  $\cpexpP=\pexpE\subst{\cpexpP}{\varX}\step{\act{}}\cpexpP'$. It is then also
  immediate that the considered derivation of
    $\pexpE\subst{\cpexpP}{\varX}\step{\act{}}\cpexpP'$
  has a derivation of $\cpexpP\step{\act{}}\cpexpP'$ as a
  subderivation.

\item
  If $\pexpE=\pref[\beta]\pexpE'$ for some $\beta\in\Acts$ and some $\varX$-closed process
  expression $\pexpE'$, then $\beta=\act$ and
  $\pexpE\step{\beta}\pexpE'$. Since
    $\pexpE\subst{\cpexpP}{\varX}=\pref[\beta](\pexpE'\subst{\cpexpP}{\varX})$,
  rule~\ref{osrule:pref} is the last rule applied in the derivation of the transition
  $\pexpE\subst{\cpexpP}{\varX}\step{\act{}}\cpexpP'$, so
  $\cpexpP'=\pexpE'\subst{\cpexpP}{\varX}$.

\item
  If $\pexpE=\rec[\varY]\pexpF$ for some process
  expression $\pexpF$ with
  $\FreeVars{\pexpF}\subseteq\{\varX,\varY\}$, then there are two
  subcases:

  On the one hand, if $\varY=\varX$, then, since $\varX$ has no free
  occurrence in $\pexpE$, we have
  $\pexpE=\pexpE\subst{\cpexpP}{\varX}\step{\act}\cpexpP'$.
  We take $\pexpE'=\cpexpP'$, and since $\pexpE'$ is 
closed we have $\pexpE'[\cpexpP/\varX] = \pexpE' = \cpexpP'$.

   On the other hand, if $\varY\neq\varX$, then
     $\pexpE\subst{\cpexpP}{\varX}
         =\rec[\varY](\pexpF\subst{\cpexpP}{\varX})$,
  and therefore the last rule applied in the considered derivation of
  the transition $\pexpE\subst{\cpexpP}{\varX}\step{\act}\cpexpP'$
  is rule~\ref{osrule:rec}.
  Consequently, the considered derivation has a proper subderivation of the transition
     $\pexpF\subst{\cpexpP}{\varX}\subst{\rec[\varY](\pexpF\subst{\cpexpP}{\varX})}{\varY}
         \step{\act}\cpexpP'$.
  Note that
     $\pexpF\subst{\cpexpP}{\varX}\subst{\rec[\varY](\pexpF\subst{\cpexpP}{\varX})}{\varY}
         =(\pexpF\subst{\rec[\varY]{\pexpF}}{\varY})\subst{\cpexpP}{\varX}$.
  Hence, by the induction hypothesis, either there exists an $\pexpE'$
  such that
    $\pexpF\subst{\rec[\varY]{\pexpF}}{\varY}\step{\act}\pexpE'$
  and
    $\cpexpP'=\pexpE'\subst{\cpexpP}{\varX}$,
  or $\varX$ is exposed
  in $\pexpF\subst{\rec[\varY]{\pexpF}}{\varY}$,
  $\cpexpP\step{\act}\cpexpP'$, and the derivation of
    $\pexpF\subst{\rec[\varY]{\pexpF}}{\varY}\subst{\cpexpP}{\varX}\step{\act{}}\cpexpP'$
  has a derivation of $\cpexpP\step{\act{}}\cpexpP'$ as a
  subderivation.
  In the first case, it follows from
    $\pexpF\subst{\rec[\varY]{\pexpF}}{\varY}\step{\act}\pexpE'$,
  by rule~\ref{osrule:rec},  that
    $\pexpE=\rec[\varY]{\pexpF}\step{\act}E'$ and
  $\cpexpP'=\pexpE'\subst{\cpexpP}{\varX}$.
  In the second case, it suffices to note that $\varX$ is exposed in
  $\pexpF$, hence also in $\pexpE$, and that a derivation of
    $\cpexpP\step{\act{}}\cpexpP'$
  appears as a subderivation of the considered derivation of
  $\pexpE\subst{\cpexpP}{\varX}\step{\act}\cpexpP'$. 

\item
  If $\pexpE=\pexpE[1]\altc\pexpE[2]$, then
    $\pexpE\subst{\cpexpP}{\varX}
       =\pexpE[1]\subst{\cpexpP}{\varX}\altc\pexpE[2]\subst{\cpexpP}{\varX}$.

  The last rule applied in the considered derivation of the transition
    $\pexpE\subst{\cpexpP}{\varX}\step{\act}\cpexpP'$
  is either rule~\ref{osrule:altcleft} or
  rule~\ref{osrule:altcright}.

  If it is rule~\ref{osrule:altcleft},
  then $\pexpE[1]\subst{\cpexpP}{\varX}\step{\act}\cpexpP'$, and since
  this transition has a derivation that is a proper subderivation of
  the considered derivation of
  $\pexpE\subst{\cpexpP}{\varX}\step{\act}\cpexpP'$, by the
  induction hypothesis it follows that either
  $\pexpE[1]\step{\act}\pexpE'$ and
    $\cpexpP'=\pexpE'\subst{\cpexpP}{\varX}$, 
  or $\varX$ is exposed in $\pexpE[1]$,
  $\cpexpP\step{\act}\cpexpP'$, and a derivation of
  $\cpexpP\step{\act{}}\cpexpP'$ appears as a subderivation the
  derivation of $\pexpE[1]\subst{\cpexpP}{\varX}\step{\act}\cpexpP'$.

  In the first case, it remains to note that then also
  $\pexpE\step{\act}\pexpE'$, and in the second case, it remains to
  note that $\varX$ is also exposed in $\pexpE$.

  If the last rule applied in the considered derivation is
  rule~\ref{osrule:altcright}, then the proof is analogous.
\qedhere
\end{itemize}
\end{proof}%
}

\begin{cor} \label{cor:liftuprec}
  Let $\pexpE$ be an $\varX$-closed process expression.
  If $\pexpE\subst{\rec[\varX]{\pexpE}}{\varX}\step{\act}\cpexpP'$
  for some (closed) process expression $\cpexpP'$, then there exists
  an $\varX$-closed process expression $\pexpE'$ such that
  $\pexpE\step{\act}\pexpE'$ and
   $\cpexpP'=\pexpE'\subst{\rec[\varX]{\pexpE}}{\varX}$.
\end{cor}
\interestingproof{%
\begin{proof}
  Consider a derivation of
    $\pexpE\subst{\rec[\varX]{\pexpE}}{\varX}\step{\act}\cpexpP'$
  that is minimal in the sense that it does not have a derivation of
    $\pexpE\subst{\rec[\varX]{\pexpE}}{\varX}\step{\act}\cpexpP'$
  as proper subderivation.
  Let $\cpexpP=\rec[\varX]{\pexpE}$.
  Since every derivation of $\cpexpP\step{\act}\cpexpP'$ has a
  derivation of $\pexpE\subst{\cpexpP}{\varX}\step{\acta{}}\cpexpP'$
  as a proper subderivation (see the operational rules, and
  rule~\ref{osrule:rec} in particular), it follows that the considered
  derivation of
    $\pexpE\subst{\rec[\varX]{\pexpE}}{\varX}\step{\act}\cpexpP'$
  does not have a subderivation of $\cpexpP\step{\act}\cpexpP'$.
  Hence, by Lemma~\ref{lem:opensteps}.\ref{liftup} there exists an
  $\varX$-closed process expression $\pexpE'$ such that
  $\pexpE\step{\act}\pexpE'$ and
  $\cpexpP'=\pexpE'\subst{\rec[\varX]{\pexpE}}{\varX}$. 
\end{proof}%
}

\begin{cor} \label{cor:liftuprec-inf}
  Let $\pexpG_0$ and $E$ be $X$-closed process expressions.
  If there is an infinite sequence of closed process expressions
  $(\cpexpP[k])_{k\in\N}$ such that
  $\pexpG_0\subst{\rec[\varX]{\pexpE}}{\varX}\mathbin{=}\cpexpP[0]$ and
  \plat{$\cpexpP[k]\step{\silent}\cpexpP[k+1]$} for all $k\in\N$, then there
  is an infinite sequence of $X$-closed process expressions
  $(G_k)_{k\in\N}$ such that $\cpexpP[k]=\pexpG_k\subst{\rec[\varX]{\pexpE}}{\varX}$
  and, for all $k\in\N$, either \plat{$G_k \step{\silent} G_{k+1}$} or $X$ is exposed in $G_k$ and \plat{$E \step{\silent} G_{k+1}$}.
\end{cor}
\interestingproof{%
\begin{proof}
We construct $(\pexpG_k)_{k\in\N}$ by induction on $k$. Suppose that $G_k$ with $\pexpG_k\subst{\rec[\varX]{\pexpE}}{\varX}\mathbin{=}\cpexpP[k]$
has already been constructed. Since \plat{$\cpexpP[k]\step{\silent}\cpexpP[k+1]$}, by
Lemma~\ref{lem:opensteps}.\ref{liftup} there are two cases:
either there is a $G_{k+1}$ with $G_k \step{\silent} G_{k+1}$
and $\cpexpP[k+1]=\pexpG_{k+1}\subst{\rec[\varX]{\pexpE}}{\varX}$, in which case we are done,
or $\varX$ is exposed in $\pexpG_k$ and $\rec[\varX]{\pexpE}\step{\silent}\cpexpP_{k+1}$.
In the latter case $\pexpE\subst{\rec[\varX]{\pexpE}}{\varX}\step{\silent}\cpexpP_{k+1}$
(see the operational rules, and rule~\ref{osrule:rec} in particular).
By Corollary~\ref{cor:liftuprec} there exists
  an $\varX$-closed process expression $\pexpG_{k+1}$ such that
  $\pexpE\step{\silent}\pexpG_{k+1}$ and
   $\cpexpP_{k+1}=\pexpG_{k+1}\subst{\rec[\varX]{\pexpE}}{\varX}$.
 \end{proof}%
 }

  Let $\pexpE$ and $\pexpE'$ be process expressions. We
  write $\pexpE\step{}\pexpE'$ if there exists an $\act\in\Acts$ such
  that $\pexpE\step{\act}\pexpE'$, and denote by $\reachsym$ the
  reflexive-transitive closure of $\step{}$. If
  $\pexpE\reach{}\pexpE'$, then we say that $\pexpE'$ is reachable
  from $\pexpE$.

\begin{prop}[{\cite[Proposition 1]{Gla93}}] \label{prop:imagefinite}
  If $\pexpE$ is a process expression, then the set of all expressions reachable from $\pexpE$
  is finite.
\end{prop}

We now characterise the relation $\rbbisimd$ on $\PEXP$ from
Definition~\ref{def:rbbisimdopen} in the same style as
Definition~\ref{def:bbisimd}, but on an enriched transition system. To
this end, we first define on $\PEXP$ a $\Vars \uplus \Acts$-labelled transition relation
${\step{}}\subseteq\PEXP\times(\Vars \uplus \Acts)\times\PEXP$ as the least ternary
relation satisfying, besides the four rules of Section~\ref{sec:bb},
also the rule 
\[  \osrule{}{\varX\step{\varX}\nil}    \label{osrule:pref}\]
for each $\varX\in\Vars$.
Intuitively, the $\Vars\uplus\Acts$-labelled transition relation
treats a process expression $\pexpE$ as the closed term obtained from
$\pexpE$ by replacing all free occurrences of the variable $\varX$ by
the closed process expression $\pref[\varX]{\nil}$ in which $\varX$ is
interpreted as an action instead of as a recursion variable.
Note that a variable $X$ is exposed in an expression $\pexpE$ according to Definition~\ref{df:exposed}
iff \plat{$\exists F.~E \step{\varX} F$}, which is the case iff \plat{$E \step{\varX}\nil$}.
Now let $\bbisimdX$ and $\rbbisimdX$ be defined exactly like $\bbisimd$ and $\rbbisimd$, but using the 
$\Vars \uplus \Acts$-labelled transition relation instead of the $\Acts$-labelled one,
and applying all definitions directly to expressions with free variables, instead of applying the
lifting of Definition~\ref{def:rbbisimdopen}.
We proceed to show that on $X$-closed process expressions $\bbisimdX$ coincides with $\bbisimd$,
and $\rbbisimdX$ with $\rbbisimd$. This characterisation, for weak and branching bisimilarity
without preservation of divergence, stems from \cite{Mil89} and \cite{Gla93}. Here we use it solely
to obtain Corollaries~\ref{cor:Xclosedtransitions} and~\ref{cor:Xclosedrootedtransition}.

\begin{lem}\label{lem:openbis}
The relation
\begin{equation*}
  \mathcal{B}=\{(\pexpE\subst{{\cpexpP}}{{\varX}},\pexpF\subst{{\cpexpP}}{{\varX}}) \mid \pexpE,\pexpF$
are $X$-closed, $\pexpE\bbisimdX\pexpF$, $P$ is closed$\}
\end{equation*}
is a branching bisimulation satisfying
(\ref{cnd:divsimshort}) of Proposition~\ref{prop:divsimalternatives}.
\end{lem}
\technicalproof{%
  \begin{proof}
It is immediate from its definition that $\mathcal{B}$ is symmetric.

We show it satisfies (T). Suppose $\pexpE,\pexpF$ are $X$-closed, $\pexpE\bbisimdX\pexpF$ and $P$ closed.
Let $\pexpE\subst{{\cpexpP}}{{\varX}}\step{\alpha} P'$ for some $\alpha\in\Acts$.
By Lemma~\ref{lem:opensteps}.\ref{liftup} either there exists an
    $\varX$-closed process expression $\pexpE'$ such that
    \plat{$\pexpE\step{\act}\pexpE'$} and
    $\cpexpP'=\pexpE'\subst{\cpexpP}{\varX}$,
    or
    $\varX$ is exposed in $\pexpE$ and \plat{$\cpexpP\step{\act{}}\cpexpP'$}.
In the first case, since $\pexpE\bbisimdX\pexpF$,
there exist process expressions $\pexpF'$ and $\pexpF''$
    such that
      \plat{$\pexpF\ssteps{}\pexpF''\step{\opt{\act}}\pexpF'$},
      $\pexpE\bbisimdX\pexpF''$
    and
      $\pexpE'\bbisimdX\pexpF'$.
By Lemma~\ref{lem:opensteps}.\ref{pushdown}
      \plat{$\pexpF\subst{{\cpexpP}}{{\varX}}\ssteps{}\pexpF''\subst{{\cpexpP}}{{\varX}}\step{\opt{\act}}\pexpF'\subst{{\cpexpP}}{{\varX}}$}.
Furthermore,
      $\pexpE\subst{{\cpexpP}}{{\varX}}\mathrel{\mathcal{B}}\pexpF''\subst{{\cpexpP}}{{\varX}}$
    and
      $P'=\pexpE'\subst{{\cpexpP}}{{\varX}}\mathrel{\mathcal{B}}\pexpF'\subst{{\cpexpP}}{{\varX}}$.
In the second case, since $\varX$ is exposed in $\pexpE$, we have
  that \plat{$\pexpE\step{\varX}\nil$} and hence, since $\pexpE\bbisimdX\pexpF$,
there exist process expressions $\pexpF'$ and $\pexpF''$
    such that
      \plat{$\pexpF\ssteps{}\pexpF''\step{X}\pexpF'$},
      $\pexpE\bbisimdX\pexpF''$
    and
      $\nil\bbisimdX\pexpF'$.
Moreover, since \plat{$\pexpF''\step{\varX}\pexpF'$}, $\varX$ is
  exposed in $\pexpF''$, so by
Lemma~\ref{lem:opensteps}.\ref{pushdown}
      \plat{$\pexpF\subst{{\cpexpP}}{{\varX}}\ssteps{}\pexpF''\subst{{\cpexpP}}{{\varX}}\step{\alpha}P'$}.
Furthermore,
      $\pexpE\subst{{\cpexpP}}{{\varX}}\mathrel{\mathcal{B}}\pexpF''\subst{{\cpexpP}}{{\varX}}$
    and
      $P'\mathrel{\mathcal{B}}P'$.

It remains to show that $\mathcal{B}$ satisfies (\ref{cnd:divsimshort}).
Suppose $\pexpE,\pexpF$ are $X$-closed, $\pexpE\bbisimdX\pexpF$ and
$P$ is closed,
and there is an infinite sequence of
     closed process expressions
         $(\cpexpP[k])_{k\in\N}$
       such that
          $\pexpE\subst{{\cpexpP}}{{\varX}}=\cpexpP[0]$ and
          \plat{$\cpexpP[k]\step{\silent}\cpexpP[k+1]$}.
By Lemma~\ref{lem:opensteps}.\ref{liftup} either there exists
an infinite sequence of $\varX$-closed process expressions
$(\pexpE_k)_{k\in\N}$ such that $\pexpE_0=\pexpE$,
    \plat{$\pexpE_k\step{\silent}\pexpE_{k+1}$} and
    $\cpexpP_{k+1}=\pexpE_{k+1}\subst{\cpexpP}{\varX}$ for all $k\in\N$,
    or
there exists
a finite sequence of $\varX$-closed process expressions
$(\pexpE_i)_{i\leq k}$ for some $k\in\N$ such that $\pexpE_0=\pexpE$,
    \plat{$\pexpE_i\step{\silent}\pexpE_{i+1}$} and
    $\cpexpP_{i+1}=\pexpE_{i+1}\subst{\cpexpP}{\varX}$ for all $i<k$,
 {$\pexpE_k\step{\varX}\nil$} and \plat{$\cpexpP\step{\silent}\cpexpP_{k+1}$}.
In the first case, since $\pexpE\bbisimdX\pexpF$, using (\ref{cnd:divsimshort}),
there exist a process expression $\pexpF'$ such that \plat{$\pexpF\step{\silent}\pexpF'$}
and $E_k \bbisimdX F'$ for some $k\in N$.
By Lemma~\ref{lem:opensteps}.\ref{pushdown}
\plat{$\pexpF\subst{\cpexpP}{\varX}\step{\silent}\pexpF'\subst{\cpexpP}{\varX}$}.
Furthermore,
      $\pexpE_k\subst{{\cpexpP}}{{\varX}}\mathrel{\mathcal{B}}\pexpF'\subst{{\cpexpP}}{{\varX}}$.
In the second case, since $\pexpE\bbisimdX\pexpF$, by induction on $i$
      there exists a sequence $\pexpF_0,\dots,\pexpF_m,\pexpF_{m+1}$
and a
      mapping $\rho:\{0,\dots,m\}\rightarrow\{0,\dots,k\}$ with $\rho(m)\mathbin{=}k$
      such that
      {$\pexpF=\pexpF_0\step{\silent}\cdots\step{\silent}\pexpF_{m}\step{X}\pexpF_{m+1}$}
 and \plat{$\pexpE[\rho(j)]\bbisimd\pexpF_{j}$} for all $j=0,\dots,m$.
If $m=0$, then $X$ is exposed in $F$, so by Lemma~\ref{lem:opensteps}.\ref{pushdown}
      \plat{$\pexpF\subst{{\cpexpP}}{{\varX}}\step{\silent}P_{k+1}$}.
Furthermore,
      $P_{k+1}\mathrel{\mathcal{B}}P_{k+1}$.
If $m\mathbin>0$, then let $F'\mathbin{=}F_1$. By Lemma~\ref{lem:opensteps}.\ref{pushdown}
\plat{$\pexpF\subst{\cpexpP}{\varX}\step{\silent}\pexpF'\subst{\cpexpP}{\varX}$}.
Furthermore, $\pexpE_{\rho(1)}\subst{{\cpexpP}}{{\varX}}\mathrel{\mathcal{B}}\pexpF'\subst{{\cpexpP}}{{\varX}}$.
\end{proof}%
}

\noindent
For every $\act\in\Acts$ and $n\in\N$, we define the closed process
expression $\act^n$ inductively by $\act^0=\nil$ and
$\act^{n+1}=\pref[\act]\act^n$. Note that, if $\act\neq\silent$, then
$\act^i\bbisimd\act^j$ implies $i=j$. Recall that we have assumed that
$\Act$ is non-empty; we now fix, for the remainder of this section, a
particular action $\acta\in\Act$.

\begin{prop}\label{prop:openbis}
Let $\pexpE$ and $\pexpF$ be $\varX$-closed process expressions.
Then $\pexpE \bbisimdX \pexpF$ iff $\pexpE \bbisimd \pexpF$,
and $\pexpE \rbbisimdX \pexpF$ iff $\pexpE \rbbisimd \pexpF$.
\end{prop}
\interestingproof{%
\begin{proof}
We need to show that \mbox{$\pexpE \bbisimdX \pexpF$} iff
$\pexpE\subst{{\cpexpP}}{{\varX}}\bbisimd\pexpF\subst{{\cpexpP}}{{\varX}}$
for each closed process expression $P$, and likewise $\pexpE \rbbisimdX \pexpF$ iff
$\pexpE\subst{{\cpexpP}}{{\varX}}\rbbisimd\pexpF\subst{{\cpexpP}}{{\varX}}$
for each closed process expression $P$.

``Only if'': Lemma~\ref{lem:openbis} immediately yields that \mbox{$\pexpE \bbisimdX \pexpF$} implies
$\pexpE\subst{{\cpexpP}}{{\varX}}\bbisimd\pexpF\subst{{\cpexpP}}{{\varX}}$
for each closed process expression $\cpexpP$.
Now let \mbox{$\pexpE \rbbisimdX \pexpF$} and \plat{$\pexpE\subst{{\cpexpP}}{{\varX}} \step{\act{}} P'$}.
By Lemma~\ref{lem:opensteps}.\ref{liftup} either there exists an
    $\varX$-closed process expression $\pexpE'$ such that
    \plat{$\pexpE\step{\act}\pexpE'$} and
    $\cpexpP'=\pexpE'\subst{\cpexpP}{\varX}$,
    or $\varX$ is exposed in $\pexpE$ and \plat{$\cpexpP\step{\act{}}\cpexpP'$}.
In the first case, since \plat{$\pexpE\rbbisimdX\pexpF$},
there exists a process expression $\pexpF'$
    such that
      \plat{$\pexpF\step{\act}\pexpF'$}
    and
      \plat{$\pexpE'\bbisimdX\pexpF'$}.
By Lemma~\ref{lem:opensteps}.\ref{pushdown}
      \plat{$\pexpF\subst{{\cpexpP}}{{\varX}}\step{\act}\pexpF'\subst{{\cpexpP}}{{\varX}}$}.
Furthermore, by Lemma~\ref{lem:openbis}
      $P'=\pexpE'\subst{{\cpexpP}}{{\varX}}\bbisimd\pexpF'\subst{{\cpexpP}}{{\varX}}$.
In the second case, since $\varX$ is exposed in $\pexpE$ we have that
\plat{$\pexpE\step{X}\nil$}, and hence, since \plat{$\pexpE\rbbisimdX\pexpF$}, there exists a process expression $\pexpF'$
    such that
      \plat{$\pexpF\step{X}\pexpF'$}.
By Lemma~\ref{lem:opensteps}.\ref{pushdown} 
      \plat{$\pexpF\subst{{\cpexpP}}{{\varX}}\step{\act}P'$}.
Furthermore, $P'\bbisimd P'$.
      The other clause follows by symmetry, thus yielding 
$\pexpE\subst{{\cpexpP}}{{\varX}}\rbbisimd\pexpF\subst{{\cpexpP}}{{\varX}}$.

``If'':
Let $\pexpE$ and $\pexpF$ be $\varX$-closed process expressions.
Since by Proposition~\ref{prop:imagefinite} the set of all process
expressions reachable from $\pexpE$ and
$\pexpF$ is finite, there exists a natural number $n\in\N$
such that for all $\pexpG$ reachable from $\pexpE$ or $\pexpF$ it holds that
$\pexpG \not\bbisimd\acta^n$, and thus
$\pexpG\subst{\acta^{n+1}}{\varX}\not\bbisimd\acta^n$.
Let
\[    \brelsym=
       \{(\pexpE',\pexpF')\mid
             \pexpE\reach{}\pexpE',\
             \pexpF\reach{}\pexpF',\
             \pexpE'\subst{\acta^{n+1}}{\varX}\bbisimd\pexpF'\subst{\acta^{n+1}}{\varX}
       \}.
\]
\emph{Claim:} The symmetric closure of $\brelsym$ is a branching bisimulation satisfying
(\ref{cnd:divsimshort}) w.r.t.\ the $\Vars \uplus \Acts$-labelled
transition relation.
\vspace{1ex}

\noindent
\emph{Proof of the claim:}
  To prove that $\brelsym{}$ satisfies condition~(\ref{cnd:stepsim}) of
  Definition~\ref{def:bbisimd}, let $\pexpE'$ and $\pexpF'$ be such
  that
     $\pexpE'\brel{}\pexpF'$,
  and suppose that
    $\pexpE'\step{\act}\pexpE''$.
  Then $\pexpE'\subst{\acta^{n+1}}{\varX}\bbisimd\pexpF'\subst{\acta^{n+1}}{\varX}$
  and, using Lemma~\ref{lem:opensteps}.\ref{pushdown},
  $\pexpE'\subst{\acta^{n+1}}{\varX}\step{\act}\pexpE''\subst{\acta^{n+1}}{\varX}$.
  Since $\pexpE'\subst{\acta^{n+1}}{\varX}\bbisimd\pexpF'\subst{\acta^{n+1}}{\varX}$
  there exist closed process expressions $\cpexpQ'''$ and $\cpexpQ''$
    such that
      $\pexpF'\subst{\acta^{n+1}}{\varX}\ssteps{}\cpexpQ''\step{\opt{\act}}\cpexpQ'''$,
      $\pexpE'\subst{\acta^{n+1}}{\varX}\bbisimd\cpexpQ''$
    and
      $\pexpE''\subst{\acta^{n+1}}{\varX}\bbisimd\cpexpQ'''$.
By Lemma~\ref{lem:opensteps}.\ref{liftup}, using that $a\neq\tau$, there exists a $\varX$-closed
process expression $\pexpF''$ such that $\pexpF' \ssteps{} \pexpF''$ and 
      $Q''=\pexpF''\subst{\acta^{n+1}}{\varX}$; moreover, either there exists an
    $\varX$-closed process expression $\pexpF'''$ such that
    \plat{$\pexpF''\step{\opt{\act}}\pexpF'''$} and
    $\cpexpQ'''=\pexpF'''\subst{\acta^{n+1}}{\varX}$,
    or $\varX$ is exposed in $\pexpF''$ and \plat{$\acta^{n+1}\step{\act{}}\cpexpQ'''$}.
In the latter case we would have $\pexpE''\subst{{\acta^{n+1}}}{{\varX}} \bbisimd Q'''=\acta^{n}$,
which is impossible by the choice of $n$. So the former case applies: we have
\plat{$\pexpF' \ssteps{} \pexpF''\step{\opt{\act}}\pexpF'''$},
$\pexpE' \mathrel{\brelsym} \pexpF''$ and
$\pexpE'' \mathrel{\brelsym} \pexpF'''$.
The case that $\pexpF'\step{\act}\pexpF''$ proceeds by symmetry, so the symmetric closure of
$\brelsym$ satisfies condition~(\ref{cnd:stepsim}).

To show that $\brelsym$ (and its symmetric closure) satisfies (\ref{cnd:divsimshort}),
let $(\pexpE[k])_{k\in\N}$ be an infinite sequence of $X$-closed process expressions
       such that
          $\pexpE[k]\step{\silent}\pexpE[k+1]$ for all $k\in\N$, and let $\pexpF[0]$ be such
  that
  $\pexpE[0]\brel{}\pexpF[0]$.
  Then $\pexpE[0]\subst{\acta^{n+1}}{\varX}\bbisimd\pexpF[0]\subst{\acta^{n+1}}{\varX}$
  and by Lemma~\ref{lem:opensteps}.\ref{pushdown}
  $\pexpE[k]\subst{\acta^{n+1}}{\varX}\step{\silent}\pexpE[k+1]\subst{\acta^{n+1}}{\varX}$ for all
  $k\mathbin\in\N$.
  Using  (\ref{cnd:divsimshort}),
there exist a process expression $\cpexpQ'$ such that \plat{$\pexpF[0]\step{\silent}\cpexpQ'$}
and $\pexpE[k]\subst{\acta^{n+1}}{\varX} \bbisimd Q'$ for some $k\in N$.
By Lemma~\ref{lem:opensteps}.\ref{liftup}, using that $a\neq\tau$, there exists a $\varX$-closed
process expression $\pexpF'$ such that $\pexpF[0] \step{\silent} \pexpF'$ and 
      $Q'=\pexpF'\subst{\acta^{n+1}}{\varX}$. Furthermore, $\pexpE[k] \mathrel{\brelsym} \pexpF'$.
\vspace{1ex}

\noindent
\emph{Application of the claim:}
Let $\pexpE\subst{{\cpexpP}}{{\varX}}\bbisimd\pexpF\subst{{\cpexpP}}{{\varX}}$
for each closed process expression $P$. Then 
$\pexpE\subst{{\acta^{n+1}}}{{\varX}}\bbisimd\pexpF\subst{{\acta^{n+1}}}{{\varX}}$.
The claim yields $\pexpE\bbisimdX\pexpF$.

Now let $\pexpE\subst{{\cpexpP}}{{\varX}}\rbbisimd\pexpF\subst{{\cpexpP}}{{\varX}}$
for each closed $P$. Then 
$\pexpE\subst{{\acta^{n+1}}}{{\varX}}\rbbisimd\pexpF\subst{{\acta^{n+1}}}{{\varX}}$.
Suppose that \plat{$\pexpE \step{\alpha} \pexpE'$} with $\alpha\in\Acts$.
Then \plat{$\pexpE\subst{{\acta^{n+1}}}{{\varX}} \step{\alpha} \pexpE'\subst{{\acta^{n+1}}}{{\varX}}$}
by Lemma~\ref{lem:opensteps}.\ref{pushdown}. So there exists a $Q'$ with
$\pexpF\subst{{\acta^{n+1}}}{{\varX}}\step{\alpha} Q'$ and
$\pexpE'\subst{{\acta^{n+1}}}{{\varX}} \bbisimd Q'$.
By Lemma~\ref{lem:opensteps}.\ref{liftup} either there exists an
    $\varX$-closed process expression $\pexpF'$ such that
    \plat{$\pexpF\step{\act}\pexpF'$} and
    $\cpexpQ'=\pexpF'\subst{\acta^{n+1}}{\varX}$,
    or $\varX$ is exposed in $\pexpF$ and \plat{$\acta^{n+1}\step{\act{}}\cpexpQ'$}.
In the latter case we would have $\pexpE'\subst{{\acta^{n+1}}}{{\varX}} \bbisimd Q'=\acta^{n}$,
which is impossible by the choice of $n$. So the former case applies, and $\pexpE' \mathrel{\brelsym} \pexpF'$.
The claim yields $\pexpE'\bbisimdX\pexpF'$.
The other clause follows by symmetry, so $\pexpE\rbbisimdX\pexpF$.
\end{proof}%
}

\noindent
The following is an immediate corollary of Propositions~\ref{prop:openbis},
\ref{prop:divsimalternatives} and~\ref{prop:bbisimdstuttering}.

\begin{cor}\label{cor:Xclosedtransitions}
  Let $\pexpE$ and $\pexpF$ be $\varX$-closed process expressions such
  that $\pexpE\bbisimd\pexpF$.
  \begin{enumerate}
  \item  \label{1} If $\pexpE\step{\act{}}\pexpE'$, then there exist
    $\varX$-closed process expressions $\pexpF[0],\dots,\pexpF[n]$ and
    $\pexpF'$ such that
    $\pexpF=\pexpF[0]\step{\silent}\cdots\step{\silent}\pexpF[n]\step{\opt{\act{}}}\pexpF'$
    such that
      $\pexpE \bbisimd\pexpF[i]$ ($0\leq i \leq n$)
    and
     $\pexpE'\bbisimd\pexpF'$.
  \item \label{2} If $\varX$ is exposed in $E$, then there exist $k\geq 0$ and
    $\varX$-closed process expressions $\pexpF[0],\dots,\pexpF[k]$
    such that
      $\pexpF=\pexpF[0]\step{\silent}\cdots\step{\silent}\pexpF[k]$,
    $\pexpE\bbisimd\pexpF[i]$ ($0\leq i \leq k$),
    and $\varX$ is exposed in $\pexpF[k]$.
  \item \label{3} If there is an infinite sequence of $\varX$-closed process
    expressions $(\pexpE[k])_{k\in\N}$ such that $\pexpE=\pexpE[0]$
    and $\pexpE[k]\step{\silent}\pexpE[k+1]$, then there exists
    an $\varX$-closed process expression $\pexpF'$ such that
    $\pexpF\step{\silent}\pexpF'$ and $\pexpE[k]\bbisimd\pexpF'$ for
    some $k\in\N$.
  \end{enumerate}
\end{cor}

\noindent
Similarly, by combining Propositions~\ref{prop:openbis}
and Definition~\ref{def:rbbisimd} we get the following corollary.

\begin{cor}\label{cor:Xclosedrootedtransition}
  Let $\pexpE$ and $\pexpF$ be $\varX$-closed process expressions such
  that $\pexpE\rbbisimd\pexpF$.
  If $\pexpE\step{\act{}}\pexpE'$, then there exists an $X$-closed
  process expression $\pexpF'$ such that
    $\pexpF\step{\act{}}\pexpF'$
  and
     $\pexpE'\bbisimd\pexpF'$.
\end{cor}

\subsection[Rooted divergence-preserving branching bisimulation up to]
  {Rooted divergence-preserving branching bisimulation up to
  \texorpdfstring{$\bbisimd$}{}} \label{subsec:rdpbbupto}

As was already illustrated by Milner \cite{Mil89h}, a suitable up-to
relation is a crucial tool in the proof that a behavioural equivalence
is compatible with the recursion construct. In \cite{Gla93}, Milner's
notion of weak bisimulation up to weak bisimilarity is adapted to
branching bisimulation up to branching bisimilarity. Here we make two
further modifications. Not only do we add a divergence condition; we
also incorporate rootedness into the relation.

\begin{defi}\label{def:upto}
  Let $\brelsym{}$ be a symmetric binary relation on $\CPEXP$, and denote by
  $\breluptobbisimdsym{}$ the relation ${\bbisimdsym\relcomp\brelsym{}\relcomp\bbisimdsym}$.
  We say that $\brelsym{}$ is a
  \emph{rooted divergence-preserving branching bisimulation up to $\bbisimd$}
  if for all $\cpexpP,\cpexpQ\in\CPEXP$ such that
  $\cpexpP\brel{}\cpexpQ$ the following three conditions are satisfied:
  \begin{enumerate}[align=right,widest=(U3)]
  \renewcommand{\labelenumi}{(\theenumi)}
  \renewcommand{\theenumi}{U\arabic{enumi}}
   \item\label{cnd:uptoshortstep}
    if
       $\cpexpP\step{\act}\cpexpP'$,
    then there exists $\cpexpQ'$ such that
    $\cpexpQ\step{\act}\cpexpQ'$
    and
       $\cpexpP'\breluptobbisimd{} Q'$.
   \item\label{cnd:uptolongstep}
    if
       $\cpexpP\ssteps{}\cpexpP''\step{\opt{\act}}\cpexpP'$,
    then there exist $\cpexpQ'$ and $\cpexpQ''$ such that
    $\cpexpQ\ssteps{}\cpexpQ''\step{\opt{\act}}\cpexpQ'$,
       $\cpexpP''\breluptobbisimd{} Q''$
    and
       $\cpexpP'\breluptobbisimd{} Q'$.
  \item\label{cnd:uptodiv}
    if there exists an infinite sequence of closed process expressions
    $(\cpexpP[k])_{k\in\N}$
       such that
          $\cpexpP=\cpexpP[0]$, and
          $\cpexpP[k]\step{\silent}\cpexpP[k+1]$ for all $k\in\N$,
    then there also exists an infinite sequence of closed process
    expressions $(\cpexpQ[\ell])_{\ell\in\N}$ and a mapping
    $\statemap{}:\N\rightarrow\N$ such that
          $\cpexpQ=\cpexpQ[0]$, and
          $\cpexpQ[\ell]\step{\silent}\cpexpQ[\ell+1]$ and
          $\cpexpP[\statemap(\ell)]\breluptobbisimd{}\cpexpQ[\ell]$ for all $\ell\in\N$.
  \end{enumerate}
\end{defi}

\begin{prop} \label{prop:uptorelatedimpliesrbbisimd}
  Let $\cpexpP$ and $\cpexpQ$ be closed process expressions and
  let $\brelsym{}$ be a rooted divergence-preserving branching bisimulation up to
  $\bbisimd$. If $\cpexpP\brel{}\cpexpQ$, then $\cpexpP\rbbisimd\cpexpQ$.
\end{prop}
\interestingproof{%
\begin{proof}
  If $\cpexpP\brel{}\cpexpQ$ and $\cpexpP\step{\act}\cpexpP'$, then
  since $\brelsym{}$ satisfies condition~(\ref{cnd:uptoshortstep}) of
  Definition~\ref{def:upto}, there exists a $\cpexpQ'$ such that
  $\cpexpQ\step{\act}\cpexpQ'$ and
  $\cpexpP'\breluptobbisimd\cpexpQ'$. Furthermore, since $\brel{}$ is
  symmetric, whenever $\cpexpP\brel{}\cpexpQ$ also
  $\cpexpQ\brel{}\cpexpP$, so if $\cpexpQ\step{\act}\cpexpQ'$, then by
  condition~(\ref{cnd:uptoshortstep}) of Definition~\ref{def:upto}
  there exists a $\cpexpP'$ such that $\cpexpP\step{\act}\cpexpP'$ and
  $\cpexpQ'\breluptobbisimd\cpexpP'$. It remains to establish that
  $\cpexpP'\bbisimd\cpexpQ'$, and for this, it suffices by
  Proposition~\ref{prop:divsimalternatives} to prove that
  $\breluptobbisimdsym$ is a branching bisimulation satisfying
  (\ref{cnd:divsimmap}).

  Note that, since $\bbisimd$ and $\brelsym{}$ are both symmetric,
  also $\breluptobbisimd$ is symmetric.

  To prove that $\breluptobbisimdsym$ satisfies (\ref{cnd:stepsim}), let
  $\cpexpP[0]$, $\cpexpP[1]$, $\cpexpQ[0]$ and $\cpexpQ[1]$ be closed
  process expressions such that
  $\cpexpP[1]\bbisimd\cpexpP[0]\brel{}\cpexpQ[0]\bbisimd\cpexpQ[1]$,
  and suppose that $\cpexpP[1]\step{\act}\cpexp{P_1'}$.
  Since $\cpexpP[1]\bbisimd\cpexpP[0]$ and $\bbisimd$ satisfies
  (\ref{cnd:stepsim}), there exist $\cpexp{P_0'}$ and $\cpexp{P_0''}$ such that
  \plat{$\cpexpP[0]\ssteps\cpexp{P_0''}\step{\opt{\act}}\cpexp{P_0'}$},
  $\cpexpP[1]\bbisimd\cpexp{P_0''}$ and
  $\cpexp{P_1'}\bbisimd\cpexp{P_0'}$.
  So it follows by condition (\ref{cnd:uptolongstep}) of
  Definition~\ref{def:upto} that there exist $\cpexp{Q_0'}$ and
  $\cpexp{Q_0''}$ such that
  \plat{$\cpexpQ[0]\ssteps\cpexp{Q_0''}\step{\opt{\act}}\cpexp{Q_0'}$},
  $\cpexp{P_0''}\breluptobbisimd\cpexp{Q_0''}$ and
  $\cpexp{P_0'}\breluptobbisimd\cpexp{Q_0'}$.
  Hence, since $\cpexpQ[0]\bbisimd\cpexpQ[1]$,  by
  Lemma~\ref{lem:longstepsim} there exist closed process expressions
  $\cpexp{Q_1'}$ and $\cpexp{Q_1''}$ such that
    \plat{$\cpexpQ[1]\ssteps\cpexp{Q_1''}\step{\opt{\act}}\cpexp{Q_1'}$},
  $\cpexp{Q_0''}\bbisimd\cpexp{Q_1''}$ and
  $\cpexp{Q_0'}\bbisimd\pexp{Q_1'}$.
  Note, moreover, that
    $\cpexpP[1]\bbisimd\cpexp{P_0''}\breluptobbisimd{}\cpexp{Q_0''}\bbisimd\cpexp{Q_1''}$
  whence $\cpexpP[1]\breluptobbisimd\cpexp{Q_1''}$,  and
    $\cpexp{P_1'}\bbisimd\cpexp{P_0'}\breluptobbisimd{}\cpexp{Q_0'}\bbisimd\cpexp{Q_1'}$
  whence $\cpexp{P_1'}\breluptobbisimd{}\cpexp{Q_1'}$.

  It remains to prove that $\breluptobbisimdsym$ satisfies
  (\ref{cnd:divsimmap}) of Proposition~\ref{prop:divsimalternatives}.
  To this end, let $\cpexpP[0]$, $\cpexpP[1]$, $\cpexpQ[0]$ and
  $\cpexpQ[1]$ be closed process expressions such that
    $\cpexpP[1]\bbisimd\cpexpP[0]\brel{}\cpexpQ[0]\bbisimd\cpexpQ[1]$,
  and suppose that there exists an infinite sequence of closed process
  expressions $(\cpexpP[1,k])_{k\in\N}$ such that
  $\cpexpP[1]=\cpexpP[1,0]$ and
  \plat{$\cpexpP[1,k]\step{\silent}\cpexpP[1,k+1]$}.
  Then, since $\cpexpP[1]\bbisimd\cpexpP[0]$, by
  Proposition~\ref{prop:divsimalternatives}, there exists an infinite
  sequence of closed process expressions $(\cpexpP[0,\ell])_{\ell\in\N}$ and
  a mapping $\statemap[\sigma_{\cpexpP}]:\N\rightarrow\N$ such that
    $\cpexpP[0]=\cpexpP[0,0]$,
    \plat{$\cpexpP[0,\ell]\step{\silent}\cpexpP[0,\ell+1]$} and
    $\cpexpP[1,{\statemap[\sigma_{\cpexpP}](\ell)}]\bbisimd\cpexpP[0,\ell]$
    for all $\ell\in\N$.
  Hence, since $\cpexpP[0]\brel\cpexpQ[0]$ and $\brelsym{}$ is a
  divergence-preserving branching bisimulation up to $\bbisimd$, there
  exists an infinite sequence of closed process expressions
    $(\cpexpQ[0,m])_{m\in\N}$
  and a mapping $\statemap[\sigma_{\cpexpP,\cpexpQ}]:\N\rightarrow\N$ such that
    $\cpexpQ[0]=\cpexpQ[0,0]$,
    \plat{$\cpexpQ[0,m]\step{\silent}\cpexpQ[0,m+1]$} and
    $\cpexpP[0,{\statemap[\sigma_{\cpexpP,\cpexpQ}](m)}]\breluptobbisimd\cpexpQ[0,m]$
    for all $m\in\N$.
  Hence, since \plat{$\cpexpQ[0]\bbisimd\cpexpQ[1]$}, by
  Proposition~\ref{prop:divsimalternatives}, there
  exists an infinite sequence of closed process expressions
    $(\cpexpQ[1,n])_{n\in\N}$
  and a mapping $\statemap[\sigma_{\cpexpQ}]:\N\rightarrow\N$ such that
    $\cpexpQ[1]=\cpexpQ[1,0]$,
    \plat{$\cpexpQ[1,n]\step{\silent}\cpexpQ[0,n+1]$} and
    $\cpexpQ[0,{\statemap[\sigma_{\cpexpQ}](n)}]\bbisimd\cpexpQ[1,n]$
    for all $n\in\N$.
  We define
  \begin{equation*}
\statemap=\statemap[\sigma_{\cpexpP}]\circ\statemap[\sigma_{\cpexpP,\cpexpQ}]\circ\statemap[\sigma_{\cpexpQ}]
  \enskip,
  \end{equation*}
  and then we have that
  $\cpexpP[1,\statemap(n)]\bbisimd\relcomp\breluptobbisimd\relcomp\bbisimd\cpexpQ[1,n]$,
  and hence $\cpexpP[1,\statemap(n)]\breluptobbisimd\cpexpQ[1,n]$ for
  all $n\in\N$.
\end{proof}%
}

To prove that $\rbbisimd$ is compatible with $\recsym[\varX]{\_}$ means
to prove that if $\pexpE\rbbisimd\pexpF$, then
$\rec[\varX]\pexpE\rbbisimd\rec[\varX]\pexpF$. We first do this in the
special case that $\pexpE$ and $\pexpF$ are $\varX$-closed. A crucial step in this
proof will
be to show that if $\pexpE\rbbisimd\pexpF$ for $\varX$-closed process
expressions $\pexpE$ and $\pexpF$, then the symmetric closure
$\urelsym[\pexpE,\pexpF]$ of the relation
  \begin{equation} \label{eq:EFupto}
      \{(\pexpG\subst{\rec[\varX]{\pexpE}}{\varX},\pexpG\subst{\rec[\varX]{\pexpF}}{\varX})
      \mid \pexpG\in\PEXP\ \text{and $\pexpG$ is $\varX$-closed} \}
  \end{equation}
is a rooted branching bisimulation up to $\bbisimd$.
The result then follows by taking $G:=X$.
Until Corollary~\ref{cor:urelcorrect} we fix $\varX$-closed process
expressions $\pexpE$ and $\pexpF$ such that $\pexpE \rbbisimd\pexpF$.

For this application of the up-to technique from Definition~\ref{def:upto},
$\breluptobbisimd{}$ could equally well have been defined as
$\brel\relcomp\bbisimdsym$. This less powerful technique is still
valid by Proposition~\ref{prop:uptorelatedimpliesrbbisimd}, yet is all we
need in Lemmas~\ref{lem:ureltransfer1}--\ref{lem:ureldiv}.

\begin{lem} \label{lem:ureltransfer1}
  For all $X$-closed process
  expressions $\pexpG$,
    if $\pexpG\subst{\rec[\varX]{\pexpE}}{\varX}\step{\act}\cpexpP$,
    then there exists a $\cpexpQ$ such that
      $\pexpG\subst{\rec[\varX]{\pexpF}}{\varX}\step{\act}\cpexpQ$
    and
      $\cpexpP \urel[\pexpE,\pexpF]\relcomp\bbisimd \cpexpQ$.
\end{lem}
\technicalproof{%
\begin{proof}
  Let $G$ be an $X$-closed process expression, and suppose that
    $\pexpG\subst{\rec[\varX]{\pexpE}}{\varX}\step{\act{}}\cpexpP$;
  we proceed by induction on a derivation of this transition.
  By Lemma~\ref{lem:opensteps}.\ref{liftup} there are two cases: either the
  transition under consideration stems directly from $\pexpG$, i.e.,
  there exists a $\pexpG'$ such that $\pexpG\step{\act{}}\pexpG'$ and
  $\cpexpP=\pexpG'\subst{\rec[\varX]{\pexpE}}{\varX}$, or $X$ is
  exposed in $G$, $\rec[\varX]{\pexpE}\step{\act}\cpexpP$ and every
  derivation of
    $\pexpG\subst{\rec[\varX]{\pexpE}}{\varX}\step{\act{}}\cpexpP$
  has a derivation of \plat{$\rec[\varX]{\pexpE}\step{\act}\cpexpP$} as a
  subderivation.

  In the first case, we have
    $\pexpG\subst{\rec[\varX]{\pexpF}}{\varX}\step{\act{}}\pexpG'\subst{\rec[\varX]{\pexpF}}{\varX}$
  and 
    $\cpexpP=
        \pexpG'\subst{\rec[\varX]{\pexpE}}{\varX}
           \urel[\pexpE,\pexpF]\pexpG'\subst{\rec[\varX]{\pexpF}}{\varX}$\
             by Lemma~\ref{lem:opensteps}.\ref{pushdown},
  so, since $\bbisimd$ is reflexive, also
   $\cpexpP\urel[\pexpE,\pexpF]\relcomp\bbisimd\pexpG'\subst{\rec[\varX]{\pexpF}}{\varX}$.

  In the second case, since the considered derivation of the transition
    $\pexpG\subst{\rec[\varX]{\pexpE}}{\varX}\step{\act}\cpexpP$
  has a derivation of
    $\rec[\varX]{\pexpE}\step{\act}\cpexpP$
  as a subderivation, and the last rule applied in this subderivation
  must be rule~\ref{osrule:rec}, it follows that the considered
  derivation of
    $\pexpG\subst{\rec[\varX]{\pexpE}}{\varX}\step{\act}\cpexpP$
  has a derivation of
     $\pexpE\subst{\rec[\varX]{\pexpE}}{\varX}\step{\act{}}\cpexpP$
  as a proper subderivation.
  So by the induction hypothesis there exists a $\cpexpQ$ such that
  $\pexpE\subst{\rec[\varX]{\pexpF}}{\varX}\step{\act{}}\cpexpQ$ and
  $\cpexpP\urel[\pexpE,\pexpF]\relcomp\bbisimd{}\cpexpQ$.
  Furthermore, since $\pexpE\rbbisimd\pexpF$, whence
  $\pexpE\subst{\rec[\varX]{\pexpF}}{\varX}\rbbisimd\pexpF\subst{\rec[\varX]{\pexpF}}{\varX}$,
  it follows that there exists an $\cpexpR$ such that
    $\pexpF\subst{\rec[\varX]{\pexpF}}{\varX}\step{\act{}}\cpexpR$
  and $\cpexpQ\bbisimd\cpexpR$.
  It follows from 
     $\pexpF\subst{\rec[\varX]{\pexpF}}{\varX}\step{\act{}}\cpexpR$
  that
     $\rec[\varX]{\pexpF}\step{\alpha}\cpexpR$.
  Since $X$ is exposed in $G$,
  Lemma~\ref{lem:opensteps}.\ref{pushdown} yields
    $\pexpG\subst{\rec[\varX]{\pexpF}}{\varX}\step{\act{}}\cpexpR$.
  From $\cpexpP\urel[\pexpE,\pexpF]\relcomp\bbisimd{}\cpexpQ$ and
  $\cpexpQ\bbisimd\cpexpR$ it follows that
  $\cpexpP\urel[\pexpE,\pexpF]\relcomp\bbisimd{}\cpexpR$.
\end{proof}%
}

\noindent
As an immediate corollary to Lemma~\ref{lem:ureltransfer1} we get that
if $\pexpE\rbbisimd\pexpF$, then $\urelsym[\pexpE,\pexpF]$ satisfies
the first condition of rooted divergence-preserving branching
bisimulations up to $\bbisimd$.

\begin{cor}\label{cor:ureluptoshortstep}
  $\urelsym[\pexpE,\pexpF]$ satisfies condition (\ref{cnd:uptoshortstep}) of
  Definition~\ref{def:upto}. 
\end{cor}

With a little more work, Lemma~\ref{lem:ureltransfer1} can also be used
to derive that $\urelsym[\pexpE,\pexpF]$ satisfies the second
condition of rooted divergence-preserving branching bisimulations up
to $\bbisimd$. To this end, we first prove the following lemma.

\begin{lem}\label{lem:ureltransfer2}
  Let
  $\cpexpP$ and $\cpexpQ$ be closed process expressions.
  If $\cpexpP\urel[\pexpE,\pexpF]\relcomp\bbisimd \cpexpQ$ and
  $\cpexpP\step{\act{}}\cpexpP'$, then there exist $\cpexpQ'$ and
  $\cpexpQ''$ such that $\cpexpQ\ssteps{}\cpexpQ''\step{\opt{\act{}}}\cpexpQ'\!$,
  $\cpexpP\urel[\pexpE,\pexpF]\relcomp\bbisimd \cpexpQ''$ and
  $\cpexpP'\urel[\pexpE,\pexpF]\relcomp\bbisimd\cpexpQ'\!$.
\end{lem}
\technicalproof{%
\begin{proof}
  Suppose that
    $\cpexpP\urel[\pexpE,\pexpF]\relcomp\bbisimd \cpexpQ$ 
  and
    $\cpexpP\step{\act{}}\cpexpP'$.
  Then there exists an $\cpexpR$ such that
    $\cpexpP\urel[\pexpE,\pexpF] \cpexpR \bbisimd \cpexpQ$,
  and according to the definition of $\urelsym[\pexpE,\pexpF]$ there
  exists an $X$-closed process expression $\pexpG$ such that either
    $\cpexpP=\pexpG\subst{\rec[\varX]{\pexpE}}{\varX}$
  and
    $\cpexpR=\pexpG\subst{\rec[\varX]{\pexpF}}{\varX}$
  or
    $\cpexpP=\pexpG\subst{\rec[\varX]{\pexpF}}{\varX}$
  and
    $\cpexpR=\pexpG\subst{\rec[\varX]{\pexpE}}{\varX}$.
  Without loss of generality we assume that
    $\cpexpP=\pexpG\subst{\rec[\varX]{\pexpE}}{\varX}$
  and
    $\cpexpR=\pexpG\subst{\rec[\varX]{\pexpF}}{\varX}$.
  By Lemma~\ref{lem:ureltransfer1}, there exists an $R'$ such that
  \plat{$\cpexpR\step{\act{}}\cpexpR'$} and
  $\cpexpP'\urel[\pexpE,\pexpF]\relcomp\bbisimd \cpexpR'$.
  Hence, since $\cpexpR\bbisimd\cpexpQ$, there exist $\cpexpQ'$ and
  $\cpexpQ''$ such that
  \plat{$\cpexpQ\ssteps{}\cpexpQ''\step{\opt{\act{}}}\cpexpQ'$},
  $\cpexpR\bbisimd\cpexpQ''$ and $\cpexpR'\bbisimd\cpexpQ'$.
  It follows that
     $\cpexpP\urel[\pexpE,\pexpF]\relcomp\bbisimd \cpexpQ''$
  and
     $\cpexpP'\urel[\pexpE,\pexpF]\relcomp\bbisimd \cpexpQ'$,
  so the proof of the lemma is complete.
\end{proof}%
}

Using that $P\urel[\pexpE,\pexpF] Q$ implies
  $P\urel[\pexpE,\pexpF]\relcomp\bbisimd Q$ by reflexivity of
  $\bbisimd$, and applying Lemma~\ref{lem:ureltransfer2} by
induction on the length of a transition sequence that gives rise to
$\cpexpP\ssteps\cpexpP''\step{\act}\cpexpP'$, it is straightforward to
establish the following corollary.

\begin{cor}\label{cor:ureluptolongstep}
  $\urelsym[\pexpE,\pexpF]$
  satisfies condition (\ref{cnd:uptolongstep}) of
  Definition~\ref{def:upto}. 
\end{cor}

It remains to establish that $\urelsym[\pexpE,\pexpF]$ satisfies the
third condition of rooted divergence-preserving branching
bisimulations up to $\bbisimd$.

\begin{lem}\label{lem:ureldiv}
  Let $\pexpG$ and $\pexpH$ be $X$-closed process
  expressions such that $\pexpG\bbisimd\pexpH$.
  If there exists an infinite sequence of closed process expressions
  $(\cpexpP[k])_{k\in\N}$ such that
  $\pexpG\subst{\rec[\varX]{\pexpE}}{\varX}\mathbin{=}\cpexpP[0]$ and
  \plat{$\cpexpP[k]\step{\silent}\cpexpP[k+1]$} for all $k\in\N$, then there
    also exists an infinite sequence of closed process expressions
  $(\cpexpQ[\ell])_{\ell\in\N}$ and a mapping
    $\statemap{}:\N\rightarrow\N$
  such that
  $\pexpH\subst{\rec[\varX]{\pexpF}}{\varX}=\cpexpQ[0]$,
  $\cpexpQ[\ell]\step{\silent}\cpexpQ[\ell+1]$, and
    $\cpexpP[\statemap(\ell)]\urel[\pexpE,\pexpF]\relcomp\bbisimd\cpexpQ[\ell]$
    for all $\ell\in\N$.
\end{lem}
\interestingproof{%
\begin{proof}
  Suppose that there exists an infinite sequence of closed process
  expressions $(\cpexpP[k])_{k\in\N}$ such that
  $\pexpG\subst{\rec[\varX]{\pexpE}}{\varX}=\cpexpP[0]$ and
  $\cpexpP[k]\step{\silent}\cpexpP[k+1]$ for all $k\in\N$.
  By Corollary~\ref{cor:liftuprec-inf} there is an infinite sequence of $X$-closed process expressions
  $(G_k)_{k\in\N}$ such that $\cpexpP[k]=\pexpG_k\subst{\rec[\varX]{\pexpE}}{\varX}$
  and either $G_k \step{\silent} G_{k+1}$ or $E \step{\silent} G_{k+1}$ for all $k\in\N$.
  We shall define simultaneously, by induction on $\ell$, an
  infinite sequence of $\varX$-closed process expressions
    $(\pexpH[\ell])_{\ell\in\N}$ with $\pexpH[0]=\pexpH$ and
    $\pexpH[\ell]\subst{\rec[\varX]{\pexpF}}{\varX}
       \step{\silent}
          \pexpH[\ell+1]\subst{\rec[\varX]{\pexpF}}{\varX}$,
  and a mapping 
    $\statemap{}:\N\rightarrow\N$,
  such that
    $\pexpG[\statemap(\ell)]\bbisimd\pexpH[\ell]$.
  This will suffice, because, for all $\ell\in\N$, defining $\cpexpQ[\ell]$ as
  $\pexpH[\ell]\subst{\rec[\varX]{\pexpF}}{\varX}$ we obtain
  $\cpexpQ[\ell]\step{\silent}\cpexpQ[\ell+1]$ and  $\cpexpP[\statemap(\ell)]=\pexpG[\statemap(\ell)]\subst{\rec[\varX]{\pexpE}}{\varX}\urel[\pexpE,\pexpF]\pexpG[\statemap(\ell)]\subst{\rec[\varX]{\pexpF}}{\varX}\bbisimd\pexpH[\ell]\subst{\rec[\varX]{\pexpF}}{\varX}=\cpexpQ[\ell]$.

  Suppose, by way of induction hypothesis, that $H_\ell$ and $\sigma(\ell)$ have been defined
  already, such that  $\pexpG[\statemap(\ell)]\bbisimd\pexpH[\ell]$.
  By Corollary~\ref{cor:liftuprec-inf} there are two cases:
    \begin{enumerate}
    \item
      $\pexpG[\statemap(\ell)+k]\step{\silent}\pexpG[\statemap(\ell)+k+1]$ for all $k\in\N$.
      Then, since $\pexpG[\statemap(\ell)]\bbisimd\pexpH[\ell]$, by
      Corollary~\ref{cor:Xclosedtransitions}.\ref{3} there exists an
      $\varX$-closed process expression $\pexpH'$ such that 
        $\pexpH[\ell]\step{\silent}\pexpH'$
      and $\pexpG[\statemap(\ell)+k]\bbisimd\pexpH'$ for some
      $k\in\N$.
      We define $\pexpH[\ell+1]=\pexpH'$ and
      $\statemap(\ell+1)=\statemap(\ell)+k$.
      Now  $\pexpH[\ell]\subst{\rec[\varX]{\pexpF}}{\varX} \step{\silent} \pexpH[\ell+1]\subst{\rec[\varX]{\pexpF}}{\varX}$ by
      Lemma~\ref{lem:opensteps}.\ref{pushdown} and $\pexpG[\statemap(\ell+1)]\bbisimd\pexpH[\ell+1]$.
    \item
      There is a $k\in\N$ such that 
     $\pexpG[\statemap(\ell)+i]\step{\silent}\pexpG[\statemap(\ell)+i+1]$ for all $i<k$,
      $X$ is exposed in $\pexpG[\statemap(\ell)+k]$ and
     \plat{$E\step{\silent}\pexpG[\statemap(\ell)+k+1]$}.
      Then, since \plat{$\pexpG[\statemap(\ell)]\bbisimd\pexpH[\ell]$}, by
      Corollary~\ref{cor:Xclosedtransitions}.\ref{1} and by induction on $i$
      there exists a sequence $\pexpH_0',\dots,\pexpH_m'$ and a
      mapping $\rho:\{0,\dots,m\}\rightarrow\{0,\dots,k\}$ with $\rho(m)\mathbin{=}k$
      such that
        $\pexpH[\ell]=\pexpH_0'\step{\silent}\cdots\step{\silent}\pexpH_{m}'$
      and
      $\pexpG[\statemap(\ell)+\rho(j)]\bbisimd\pexpH_{j}'$.
      Using Corollary~\ref{cor:Xclosedtransitions}.\ref{2}, we may furthermore assume that $X$ is
      exposed in $H'_m$.

      If $m>0$, then we define
      $\pexpH[\ell+1]=\pexpH_1'$ and
      $\sigma(\ell+1)=\sigma(\ell)+\rho(1)$.
      Now  $\pexpH[\ell]\subst{\rec[\varX]{\pexpF}}{\varX} \step{\silent} \pexpH[\ell+1]\subst{\rec[\varX]{\pexpF}}{\varX}$ by
      Lemma~\ref{lem:opensteps}.\ref{pushdown} and $\pexpG[\statemap(\ell+1)]\bbisimd\pexpH[\ell+1]$.

      So it remains to consider the case that $m=0$.
      Since $\pexpE\rbbisimd\pexpF$, there exists, by
      Corollary~\ref{cor:Xclosedrootedtransition}, an $\varX$-closed
      process expression $\pexpF'$ such that
      \plat{$\pexpF\step{\silent}\pexpF'$} and $\pexpG[\statemap(\ell)+k+1]\bbisimd\pexpF'$.
      We now define $\pexpH[\ell+1]=\pexpF'$ and
      $\statemap(\ell+1)=\statemap(\ell)+k+1$.
      We then have that
      $\pexpG[\statemap(\ell+1)]=\pexpG[\statemap(\ell)+k+1]\bbisimd\pexpH[\ell+1]$,
      and
      \plat{$\pexpF\subst{\rec[\varX]{\pexpF}}{\varX}\step{\silent}\pexpH[\ell+1]\subst{\rec[\varX]{\pexpF}}{\varX}$}
      by Lemma~\ref{lem:opensteps}.\ref{pushdown}.
      So \plat{$\rec[\varX]{\pexpF}\step{\silent}\pexpH[\ell+1]\subst{\rec[\varX]{\pexpF}}{\varX}$}
      by rule~\ref{osrule:rec}, and Lemma~\ref{lem:opensteps}.\ref{pushdown} yields
      $\pexpH[\ell]\subst{\rec[\varX]{\pexpF}}{\varX}\step{\silent}\pexpH[\ell+1]\subst{\rec[\varX]{\pexpF}}{\varX}$,
      using that $X$ is exposed in $\pexpH[\ell]$.
\qedhere
    \end{enumerate}
\end{proof}%
}

\noindent
From Lemma~\ref{lem:ureldiv} with $\pexpG=\pexpH$ we immediately get the following corollary.

\begin{cor}\label{cor:ureldiv}
  $\urelsym[\pexpE,\pexpF]$
  satisfies condition (\ref{cnd:uptodiv}) of
  Definition~\ref{def:upto}. 
\end{cor}

The relation $\urelsym[\pexpE,\pexpF]$ is symmetric by definition and
we have now also proved that it satisfies conditions
(\ref{cnd:uptoshortstep}), (\ref{cnd:uptolongstep}) and
(\ref{cnd:uptodiv}), so we have established the following result.
\begin{cor} \label{cor:urelcorrect}
  $\urelsym[\pexpE,\pexpF]$ is a rooted divergence-preserving branching bisimulation up to $\bbisimd$.
\end{cor}

\subsection[The main results]{The main results} \label{subsec:main}
\noindent
We can now establish that $\rbbisimd$ is compatible with
$\prefsym[\act{}]$, $\recsym[\varX]{\_}$ and~$\altc$.

\begin{prop}\label{prop:compatibility}
  If $\pexpE\rbbisimd\pexpF$, then
  $\pref[\act]{\pexpE}\rbbisimd\pref[\act]{\pexpF}$ for all
  $\act\in\Acts$, 
  $\pexpE\altc\pexpH\rbbisimd\pexpF\altc\pexpH$ and
  $\pexpH\altc\pexpE\rbbisimd\pexpH\altc\pexpF$ for all process
  expressions $\pexpH$, and
  $\rec[\varX]{\pexpE}\rbbisimd\rec[\varX]{\pexpF}$.
\end{prop}
\technicalproof{%
\begin{proof}
  To prove that $\rbbisimd$ is compatible with $\prefsym[\act{}]$ and
  $\altc$ is straightforward. (First, establish the property for
  closed terms, and then use that substitution distributes over
  $\prefsym[\act{}]$ and $\altc$.)

  It remains to prove that $\rbbisimd$ is compatible with
  $\recsym[\varX]{\_}$, i.e., that $\pexpE\rbbisimd\pexpF$ implies
  $\rec[\varX]{\pexpE}\rbbisimd\rec[\varX]{\pexpF}$.
  Note that in the special case that $\pexpE$ and $\pexpF$ are
  $\varX$-closed this immediately follows from
  Corollary~\ref{cor:urelcorrect} and
  Proposition~\ref{prop:uptorelatedimpliesrbbisimd}.
  Now, for the general case, let $\pexpE$ and $\pexpF$ be process
  expressions and suppose that $\pexpE\rbbisimd\pexpF$.
  Let $\varX,\vec{\varY}$ be a sequence of variables that at
  least includes the variables with a free occurrence in $\pexpE$ or
  $\pexpF$, and such that $\varX$ does not occur in
  \plat{$\vec{\varY}$}.
  Then, according to the definition of $\rbbisimd$ on
  process expressions with free variables
  (Definition~\ref{def:rbbisimdopen}), we have that, for every closed
  process expression $\cpexpP$ and for every sequence of closed process
  expressions $\vec{\cpexpP}$ of the same length as $\vec{Y}$,
    $\pexpE\subst{\cpexpP,\vec{\cpexpP}}{\varX,\vec{\varY}}
        \rbbisimd
      \pexpF\subst{\cpexpP,\vec{\cpexpP}}{\varX,\vec{\varY}}$.
  So, clearly, also
    $\pexpE\subst{\vec{\cpexpP}}{\vec{\varY}}
        \rbbisimd
      \pexpF\subst{\vec{\cpexpP}}{\vec{\varY}}$,
   and since $\pexpE\subst{\vec{\cpexpP}}{\vec{\varY}}$ and
   $\pexpF\subst{\vec{\cpexpP}}{\vec{\varY}}$ are
   $\varX$-closed, it follows that
     $\rec[\varX]{\pexpE}\subst{\vec{\cpexpP}}{\vec{\varY}}
        \rbbisimd
      \rec[\varX]{\pexpF}\subst{\vec{\cpexpP}}{\vec{\varY}}$.
   Since $\varX$ is not among the $\vec{\varY}$, we may conclude that
     $(\rec[\varX]{\pexpE})\subst{\vec{\cpexpP}}{\vec{\varY}}
        \rbbisimd
      (\rec[\varX]{\pexpF})\subst{\vec{\cpexpP}}{\vec{\varY}}$
   for every sequence of closed process expressions $\vec{\cpexpP}$ of
   the same length as $\vec{\varY}$,
   and hence
     $\rec[\varX]{\pexpE}\rbbisimd\rec[\varX]{\pexpF}$.
\end{proof}%
}

\noindent
We have now obtained our main result that $\rbbisimd$ is a
congruence. In fact, it is the coarsest contained in $\bbisimd$.
\begin{thm}
  The relation $\rbbisimd$ is the coarsest congruence contained in $\bbisimd$.
\end{thm}
\technicalproof{%
\begin{proof}
  By Propositions~\ref{prop:equivalence} and \ref{prop:compatibility}, the relation $\rbbisimd$ is a
  congruence.
  To prove that it is the coarsest, it suffices to prove that for every relation
  $\mathcal{R}\subseteq{\bbisimd}$ that is compatible with $\altc$ we
  have that $\mathcal{R}\subseteq{\rbbisimd}$.
  Let $\cpexpP$ and $\cpexpQ$ be closed process expressions, and
  suppose that $\cpexpP\brel\cpexpQ$.

  Since by Proposition~\ref{prop:imagefinite} the set of
  closed process expressions reachable from $\cpexpP$ and $\cpexpQ$ is
  finite and $\Act$ is non-empty, there exists a natural number
  $n\in\N$ such that for all $\cpexpR$ reachable from $\cpexpP$ or
  $\cpexpQ$ it holds that $\cpexpR\not\bbisimd\acta^n$. This implies
  that for all $\cpexpR'$ reachable from $\cpexpP$ or $\cpexpQ$ it
  holds that $\cpexpR'\not\bbisimd\cpexpP\altc\acta^{n+1}$ and
  $\cpexpR'\not\bbisimd\cpexpQ\altc\acta^{n+1}$; for suppose
    that, e.g., there exists $\cpexpR'$ reachable from $\cpexpP$ or
    $\cpexpQ$ such that $\cpexpR'\bbisimd\cpexpP\altc\acta^{n+1}$,
    then, since $\cpexpP\altc\acta^{n+1}\step{\acta}\acta^n$, we have
    that $\acta^n$ is reachable from $\cpexpP$ or $\cpexpQ$.

  Since $\mathcal{R}$ is compatible with $\altc$, we have that
  $\cpexpP\altc \acta^{n+1} \brel{} \cpexpQ\altc \acta^{n+1}$, and hence
  $\cpexpP\altc \acta^{n+1} \bbisimd \cpexpQ\altc \acta^{n+1}$.
  To prove (\ref{cnd:rootsiml}), suppose that
  $\cpexpP\step{\act}\cpexpP'$. Then
  $\cpexpP\altc\acta^{n+1}\step{\act}\cpexpP'$, so by
  Lemma~\ref{lem:longstepsim} there exist closed process expressions $\cpexpQ'$ and $\cpexpQ''$
  such that $\cpexpQ\altc\acta^{n+1}\ssteps\cpexpQ''\step{\opt{\act}}\cpexpQ'$,
  $\cpexpP\altc\acta^{n+1}\bbisimd\cpexpQ''$ and $\cpexpP'\bbisimd\cpexpQ'$. Since
  $\acta\neq\silent$, we have that $\cpexpQ''=\cpexpQ+\acta^{n+1}$,
  for otherwise $\cpexpQ''$ is reachable from $\cpexpQ$ and
  $\cpexpQ''\bbisimd\cpexpP\altc\acta^{n+1}$. Moreover,
  $\cpexpQ''\step{\act}\cpexpQ'$, for otherwise
  $\cpexpP'\bbisimd\cpexpQ'=\cpexpQ''=\cpexpQ\altc\acta^{n+1}$.
  Condition (\ref{cnd:rootsimr}) follows by symmetry.  
\end{proof}%
}

\bibliographystyle{alpha}
\def\sortunder#1{}
\bibliography{paper}

\end{document}